\documentclass[a4paper,fleqn,usenatbib]{mnras}

\usepackage{newtxtext,newtxmath}

\usepackage[T1]{fontenc}
\usepackage{ae,aecompl}
\usepackage{dsfont}
\usepackage{color}
\usepackage{microtype}

\usepackage{cite}
\usepackage{rotating, tabularx}
\usepackage{bm}
\usepackage{longtable}
\usepackage{graphicx}	
\usepackage{amsmath}	
\usepackage{amssymb}	

\newcommand{\sigmalsf}{$\sigma_{\text{LSF}}$}
\newcommand{\sigmavlsf}{$\sigma_{v, \text{LSF}}$}
\newcommand{\sigmavlos}{$\sigma_{v, \text{LoS}}$}
\newcommand{\sigmavz}{$\sigma_{v, z}$}
\newcommand{\sigmav}{$\sigma_{v}$}

\newcommand{\sfrdensity}{$\Sigma_\text{SFR}$}

\newcommand{\kms}{km s$^{-1}$}
\newcommand{\smassyr}{M$_\odot$ yr$^{-1}$}
\newcommand{\pmstar}{$\langle p_*/m_* \rangle$}
\newcommand{\eff}{$\epsilon_\text{ff}$}

\newcommand{\blobby}{\textsc{Blobby3D}}
\newcommand{\bbarolo}{\textsc{$^\mathrm{3D}$Barolo}}
\newcommand{\dnest}{\textsc{DNest4}}

\DeclareRobustCommand{\ion}[2]{%
\relax\ifmmode
\ifx\testbx\f@series
{\mathbf{#1\,\mathsc{#2}}}\else
{\mathrm{#1\,\mathsc{#2}}}\fi
\else\textup{#1\,{\mdseries\textsc{#2}}}%
\fi}

\title[Drivers of gas turbulence]{The SAMI Galaxy Survey: Gas velocity dispersions in low-$z$ star-forming galaxies and the drivers of turbulence}

\author[Varidel et al.]{Mathew R. Varidel$^{1,2}$\thanks{E-mail: mathew.varidel@sydney.edu.au},
Scott M. Croom$^{1,2}$,
Geraint F. Lewis$^{1}$,
Deanne B. Fisher$^{2,3}$,
\newauthor
Karl Glazebrook$^{2,3}$,
Barbara Catinella$^{2,4}$,
Luca Cortese$^{2,4}$,
Mark R. Krumholz$^{2,5,6,7}$,
\newauthor
Joss Bland-Hawthorn$^{1,2}$,
Julia J. Bryant$^{1,2,8}$,
Brent Groves$^{2,4,5}$,
Sarah Brough$^{2,9}$,
\newauthor
Christoph Federrath$^{5}$,
Jon S. Lawrence$^{10}$,
Nuria P. Lorente$^{11}$,
Matt S. Owers$^{12, 13}$,
\newauthor
Samuel N. Richards$^{14}$,
\'Angel R. L\'opez-S\'anchez$^{8,12}$,
Sarah M. Sweet$^{2,15}$,
\newauthor
Jesse van de Sande$^{1,2}$, and
Sam P. Vaughan$^{1,2}$
\\
$^{1}$Sydney Institute for Astronomy (SIfA), School of Physics, A28, The University of Sydney, NSW 2006, Australia\\
$^{2}$ARC Centre of Excellence for All Sky Astrophysics in 3 Dimensions (ASTRO 3D)\\
$^{3}$Centre for Astrophysics and Supercomputing, Swinburne University of Technology, PO Box 218, Hawthorn, VIC 3122, Australia\\
$^{4}$International Centre for Radio Astronomy Research, University of Western Australia, 35 Stirling Highway, Crawley WA 6009, Australia\\
$^{5}$Research School of Astronomy and Astrophysics, Australian National University, Canberra, ACT 2611, Australia\\
$^{6}$Universit\"at Heidelberg, Zentrum f\"ur Astronomie, Institut f\"ur Theoretische Astrophysik, 69120 Heidelberg, Germany\\
$^{7}$Max Planck Institute for Astronomy, K\"onigstuhl 17, 69117 Heidelberg, Germany\\
$^{8}$Australian Astronomical Optics, AAO-USydney, School of Physics, University of Sydney, NSW 2006, Australia\\
$^{9}$School of Physics, University of New South Wales, NSW 2052, Australia\\
$^{10}$Australian Astronomical Optics, Macquarie University, 105 Delhi Rd, North Ryde, NSW 2113, Australia\\
$^{11}$Faculty of Science \& Engineering, Macquarie University. 105 Delhi Rd, North Ryde, NSW 2113, Australia\\
$^{12}$Department of Physics and Astronomy, Macquarie University, NSW 2109, Australia\\
$^{13}$Astronomy, Astrophysics and Astrophotonics Research Centre, Macquarie University, Sydney, NSW 2109, Australia\\
$^{14}$SOFIA Science Center, USRA, NASA Ames Research Center, Building N232, M/S 232-12, P.O. Box 1, Moffett Field, CA 94035-0001, USA\\
$^{15}$School of Mathematics and Physics, University of Queensland, Brisbane, QLD 4072, Australia\\
}

\date{Accepted XXX. Received YYY; in original form ZZZ}

\pubyear{2020}

\begin{document}
\label{firstpage}
\pagerange{\pageref{firstpage}--\pageref{lastpage}}
\maketitle

\begin{abstract}
We infer the intrinsic ionised gas kinematics for 383 star-forming galaxies across a range of integrated star-formation rates (SFR $\in [10^{-3}, 10^2]$ \smassyr{}) at $z \lesssim 0.1$ using a consistent 3D forward-modelling technique. The total sample is a combination of galaxies from the SAMI Galaxy Survey and DYNAMO survey. For typical low-$z$ galaxies taken from the SAMI Galaxy Survey, we find the vertical velocity dispersion (\sigmavz{}) to be positively correlated with measures of star-formation rate, stellar mass, \ion{H}{i} gas mass, and rotational velocity. The greatest correlation is with star-formation rate surface density (\sfrdensity{}). Using the total sample, we find \sigmavz{} increases slowly as a function of integrated star-formation rate in the range \mbox{SFR $\in$ [$10^{-3}$, 1] \smassyr{}} from $17\pm3$ \kms{} to $24\pm5$ \kms{} followed by a steeper increase up to \sigmavz{} $\sim 80$ \kms{} for \mbox{SFR $\gtrsim 1$  \smassyr{}}. This is consistent with recent theoretical models that suggest a \sigmavz{} floor driven by star-formation feedback processes with an upturn in \sigmavz{} at higher SFR driven by gravitational transport of gas through the disc.

\end{abstract}




\clearpage
\section{Introduction}
\label{sec:intro}

Galaxies at $z > 1$ typically have velocity dispersions greater than nearby galaxies \citep{Kassin2012,Wisnioski2015,Johnson2018,Ubler2019}. While observations of galaxies at $z > 1$ reveal a significant proportion of galaxies with velocity dispersions in the range 50 -- 100 \kms{} \citep[e.g.][]{Genzel2006,Law2007,ForsterSchreiber2009,Law2009,Epinat2010,Jones2010,LemoineBusserolle2010}, nearby galaxies typically have velocity dispersions of < 50 \kms{} \citep{Epinat2008,Moiseev2015,Varidel2016,Yu2019}. Although this has been observed, the process by which galaxies settle to lower velocity dispersions across epochs is not well understood.

Another important observation is that galaxies at all epochs exhibit velocity dispersions that are greater than expected by the thermal contribution of the gas alone. In the case of ionised gas measured using the H$\alpha$ emission line, the characteristic temperature of 10$^4$ K corresponds to an expected velocity dispersion of $\sim$9 \kms{} \citep{Glazebrook2013}. Galaxies have velocity dispersions > 9 \kms{} at all epochs.

Studies suggest that turbulent motions above the thermal contribution dissipate on timescales of the order of the flow crossing time \citep{MacLow1998,Stone1998,MacLow1999}. The crossing time for a galaxy with Toomre stability \citep{Toomre1964} of $Q \sim 1$ will be of order the dynamical time, which is typically $\mathcal{O}$(100 Myr) \citep{Krumholz2018}. If the turbulent motions are on the scale of Giant Molecular Clouds (GMCs), it will decay on $\mathcal{O}$(< 10 Myr). Therefore, we should rarely see galaxies with velocity dispersions greater than the thermal contribution, unless there is an ongoing driving mechanism to sustain the observed gas turbulence.

Numerous energetic sources have been proposed to contribute to the non-thermal turbulence observed in galaxies. These drivers can typically be split into star-formation feedback processes \citep{Norman1996,MacLow2004,Krumholz2009,Murray2010}, gravitational transport of gas onto \citep{Elmegreen2010,Hopkins2013} or through \citep{Krumholz2010} the disc, dynamical drivers such as shear and differential rotations across the disc \citep{Federrath2016,Federrath2017IAUS}, or interactions between galaxy components \citep[e.g.][]{Dobbs2007,Dekel2009a,Ceverino2010,Aumer2010,Oliva-Altamirano2018}. In this paper, we will be focusing primarily on differentiating star-formation feedback processes and gravitational transport of gas through the disc due to the clear predictions that have been made in the integrated star-formation rate (SFR) and global velocity dispersion ($\sigma_v$) plane \citep{Krumholz2016,Krumholz2018}.

Star-formation feedback is thought to be dominated by the energy imparted by supernovae \citep{Norman1996,MacLow2004}. However, other drivers such as stellar winds, expansion of \ion{H}{ii} regions \citep{Chu1994,Matzner2002}, and radiation pressure in high density star clusters \mbox{\citep{Krumholz2009,Murray2010}} will also inject momentum into the interstellar medium. Observational evidence for star-formation feedback as the primary driver of gas turbulence has been argued by observing that SFR is correlated with $\sigma_v$. The SFR -- $\sigma_v$ correlation has been shown both within a single sample at constant redshift \citep{Green2010,Green2014,Moiseev2015,Yu2019} and by combining multiple samples across epochs \citep{Green2010,Green2014}.

Assuming that star-formation feedback processes are a significant driver of the turbulence, it would be natural to expect a relation between local star-formation rate surface density ($\Sigma_\text{SFR}$) and local velocity dispersion. There are conflicting results in the literature regarding the relationship between these local quantities. Some studies have found a significant relationship \citep{Lehnert2009,Lehnert2013}, whereas others have found the localised relationship to be weak  \citep{Genzel2011,Varidel2016,Zhou2017,Ubler2019}.


Furthermore, the physical mechanism for an energetic source to account for velocity dispersions due to star-formation feedback of several tens of \kms{} is not well established. Constructing equilibrium solutions between gravitational infall of the disc \mbox{supported} by outward pressure solely by supernovae leads to $\sigma_v \lesssim 25$ \kms{} with little variation as a function of SFR \citep{Ostriker2011,Krumholz2018}. An alternative approach that can \mbox{account} for increased turbulence is to assume that the star-formation efficiency per free-fall time (\eff{}) changes as a function of galaxy properties, thus changing the energetic input from star-formation feedback processes \citep{Faucher-Giguere2013}. However, \mbox{numerous} observations suggest that \eff{} is approximately constant across a wide range of galaxy properties \citep{Krumholz2007,Krumholz2012,Federrath2013,Salim2015,Krumholz2019}.

An alternative set of driving mechanisms are due to gravitational effects. This includes the initial gravitationally unstable formation of the disc \citep{Aumer2010}, that can account for short-lived supersonic turbulence on the order of the disc formation time, $\mathcal{O}$(100 Myr). It is thought that the supersonic turbulence that is initially set at disc formation can be maintained by the gravitational transport of gas through the disc \citep{Krumholz2010}. \citet{Krumholz2016} also argued that the gravitational transport model predicts an increase in velocity dispersion at increased SFR that is more consistent with the data than models assuming star-formation feedback processes.

A further complication involved in inferring the ongoing drivers of turbulence across epochs is the effects of the spectral and spatial resolution on the observed velocity dispersion. The spectral resolution broadens the observed emission line often on order of the intrinsic velocity dispersion. This is typically accounted for by convolving the modelled emission line profile by the known Line-Spread Function (LSF) while fitting to the data \citep[e.g.][]{ForsterSchreiber2009,Davies2011,Green2014,Varidel2019}. This is a reasonable approximation as long as the model assumptions regarding the LSF are well known.

The spatial resolution is more difficult to account for as it acts to blur the emission line flux spatially per spectral slice. The observed velocity dispersion is then a complex function of the intrinsic flux distribution, line of sight (LoS) velocity profile, and LoS velocity dispersion profile. This effect is usually referred to as beam smearing.

In general, beam smearing acts to increase the observed velocity dispersion particularly where the velocity gradient is steepest \citep{Davies2011,Glazebrook2013}, and in detail can result in spurious substructure in the velocity dispersion profile \citep{Varidel2019}. Furthermore, beam smearing could result in spurious correlations such as the SFR -- $\sigma_v$ correlation, as SFR is related to the mass which shapes the gravitational potential, and thus increases the velocity gradient at the centre of galaxies with higher SFR. Similarly, the width of the Point-Spread Function (PSF) relative to the galaxy size increases for increasing $z$, thus resulting in higher observed velocity dispersions if beam smearing is not corrected for appropriately.

The SFR -- $\sigma_v$ relation has been used to distinguish between the different energetic sources of turbulence \citep{Krumholz2016,Krumholz2018}. However, comparisons between theoretical models and observations have typically been performed by combining several studies with different redshift ranges and beam smearing corrections. In this paper, we improve comparisons of the observed velocity dispersion to theoretical models by studying a sample of nearby galaxies using a single  technique to mitigate the effects of beam smearing. The data encompasses a wide range of \mbox{SFR $\in$ [10$^{-3}$, 10$^2$] \smassyr{}} of local galaxies at $z \lesssim 0.1$. The combined sample is comprised of observations from the SAMI Galaxy Survey Data Release Two  \citep[SAMI Galaxy Survey DR2,][]{Croom2012,Scott2018} and the DYNAMO survey \citep{Green2014}. We use a consistent disc-fitting routine referred to as \blobby{} \citep{Varidel2019}, for all the galaxy gas kinematic modelling in this paper. \blobby{} is a disc fitting code that constructs a regularly rotating thin-disc galaxy model in 3D (position -- position -- wavelength space) that is then convolved by the PSF and LSF prior to comparing the model to the data. In that way it can account for the effect of beam smearing when inferring the velocity dispersion of the galaxy.

The outline of this paper is as follows. In Section \ref{sec:data} we describe the SAMI Galaxy Survey and DYNAMO surveys, as well as our sample selection criteria. In Section \ref{sec:methods} we outline the methods used to measure the key gas kinematic properties. In Section \ref{sec:results}, we will discuss our results. In Section \ref{sec:vdisp_drivers} we compare our results to theoretical models of the drivers for turbulence. We summarise our conclusions in Section \ref{sec:conclusions}. Throughout this paper we assume the concordance cosmology   \citep[$\Omega_\Lambda$ = 0.7, $\Omega_m$ = 0.3, $H_0$ = 70 km s$^{-1}$ Mpc$^{-1}$;][]{Hinshaw2009} and a \citet{Chabrier2003} Initial Mass Function (IMF).

\section{Data selection}
\label{sec:data}

\subsection{The SAMI Galaxy Survey}
\label{subsec:SAMI}

The SAMI Galaxy Survey was conducted with the Sydney-AAO Multi-object Integral field Spectrograph \citep[SAMI,][]{Croom2012}. SAMI was mounted at the Anglo-Australian Telescope (AAT), that provided a 1 degree diameter Field-of-View (FoV). SAMI used 13 fused fibre bundles, known as Hexabundles \citep{Bland-Hawthorn2011,Bryant2014}, with a 75\% fill factor. Each bundle contains 61 fibres of 1.6$''$ diameter, resulting in an approximately 15$''$ diameter FoV. The IFUs as well as 26 sky fibres were attached to pre-drilled plates using magnetic connectors. SAMI fibres were fed to the double-beam AAOmega spectrograph \citep{Sharp2006}. The 580V grating at 3750--5750 \AA{} provides a resolution of \mbox{$R = 1808$} (\mbox{$\sigma = 70.4$ km s$^{-1}$} at 4800 \AA{}) and the 1000R grating from \mbox{6300--7400 \AA{}} providing a resolution of $R = 4304$ (\mbox{$\sigma = 29.6$ km s$^{-1}$} at 6850 \AA{}) \citep{Scott2018}.

During the survey, observations of over 3000 galaxies were obtained. Target selection for the SAMI Galaxy Survey are provided in \citet{Bryant2015}. The redshift range for the observed galaxies was $0.004 < z < 0.113$ and a stellar mass range of $7.5 < \log(M_*/M_\odot) < 11.6$. The Full-Width Half-Maximum (FWHM) of the seeing distribution was $1.10'' < \text{FWHM}_\text{PSF} < 3.27''$. Relevant data used for the analysis in this paper are from the SAMI Galaxy Survey DR2 \citep{Scott2018}. This includes the aperture spectra, emission line products \citep{Green2018}, data cubes \citep{Sharp2015}, and input catalogue \citep{Bryant2015}.

\subsection{Sample selection from the SAMI Galaxy Survey}
\label{subsec:Sample}

Our aim was to select galaxies on the star-forming main sequence within the SAMI Galaxy Survey. As such, we performed the following selection criteria cuts to the sample from the SAMI Galaxy Survey DR2 \citep{Scott2018}.

Star-forming galaxies are selected by applying a cutoff integrated H$\alpha$ equivalent width of $EW > 3$ \AA{} \citep{CidFernandes2011}. The equivalent width is calculated as the total H$\alpha$ flux compared to the total continuum flux across the SAMI FoV. The continuum flux in the region around H$\alpha$ is estimated by calculating the mean continuum in the wavelength range [6500, 6540] \AA{}. The integrated H$\alpha$ flux estimates is sourced from the SAMI Galaxy Survey DR2 emission line data products.

We remove galaxies with ionised emission from non star-forming sources such as Active Galactic Nuclei (AGN) and Low-Ionisation Nuclear Emission-line Regions (LINERs). To implement this criteria, we remove galaxies where the AGN classification criteria proposed by \citet{Kauffmann2003} is met,
\begin{equation}
    \log (
        [\text{\ion{O}{iii}}] / \text{H} \beta
        ) > 
        \frac{0.61}{\log([\text{\ion{N}{ii}}]/\text{H}\alpha) - 0.05} + 1.3.
    \label{eq:kauffmanagn}
\end{equation}
[\ion{O}{iii}] and [\ion{N}{ii}] represent the emission line fluxes at 5007 \AA{} and 6583 \AA{}, respectively. The line fluxes are estimated for the central region of the galaxy where AGN and LINER contamination should be greatest, using the 1.4$''$ aperture spectra from the SAMI Galaxy Survey DR2.

We retain galaxies that are face-on up to $e = 1 - b/a = 0.5$ (0$^{\circ}$ < $i$ < 60$^{\circ}$, assuming a thin disc). We avoid galaxies observed at high inclination as the intrinsic velocity dispersion is more difficult to constrain due to beam smearing. Plus galaxies are optically thick such that edge-on observations limit the ability to observe the integrated LoS from the entire galaxy. Furthermore, a thin disc model is assumed in \blobby{}, such that the galaxies will not be well modelled when observed close to edge-on.

We apply the following signal-to-noise cut on the spaxels in the data. We first apply a mask to spaxels with H$\alpha$ flux signal-to-noise < 3. Spatially resolved H$\alpha$ flux and it's error are obtained from the SAMI Galaxy Survey DR2 pipeline. We then construct groups of unmasked spaxels that are adjacent and meet the signal-to-noise criteria. The largest unmasked group is retained, whereas the remaining spaxels are masked. We retain galaxies that had at least 300 unmasked spaxels.

The above masking routine only finds the largest group of spaxels, which in principle could reject clumpy flux profiles. In practice, the effect of removing H$\alpha$ clumps originating from the galaxy was negligible. Instead, it primarily removed spurious spaxels that were reported to have high signal-to-noise, yet by eye did not appear to be legitimate detections of flux originating from the galaxy.

We also remove mergers or galaxies with clearly disturbed gas kinematics from our final sample. Potential mergers were determined by eye from observations of the gas kinematic maps. 9 galaxies were removed from our final sample due to this criteria.

\begin{figure*}
    \begin{center}
        \includegraphics[
            width=\textwidth,
            keepaspectratio=true,
            trim=10mm 20mm 20mm 10mm,
            clip=true]{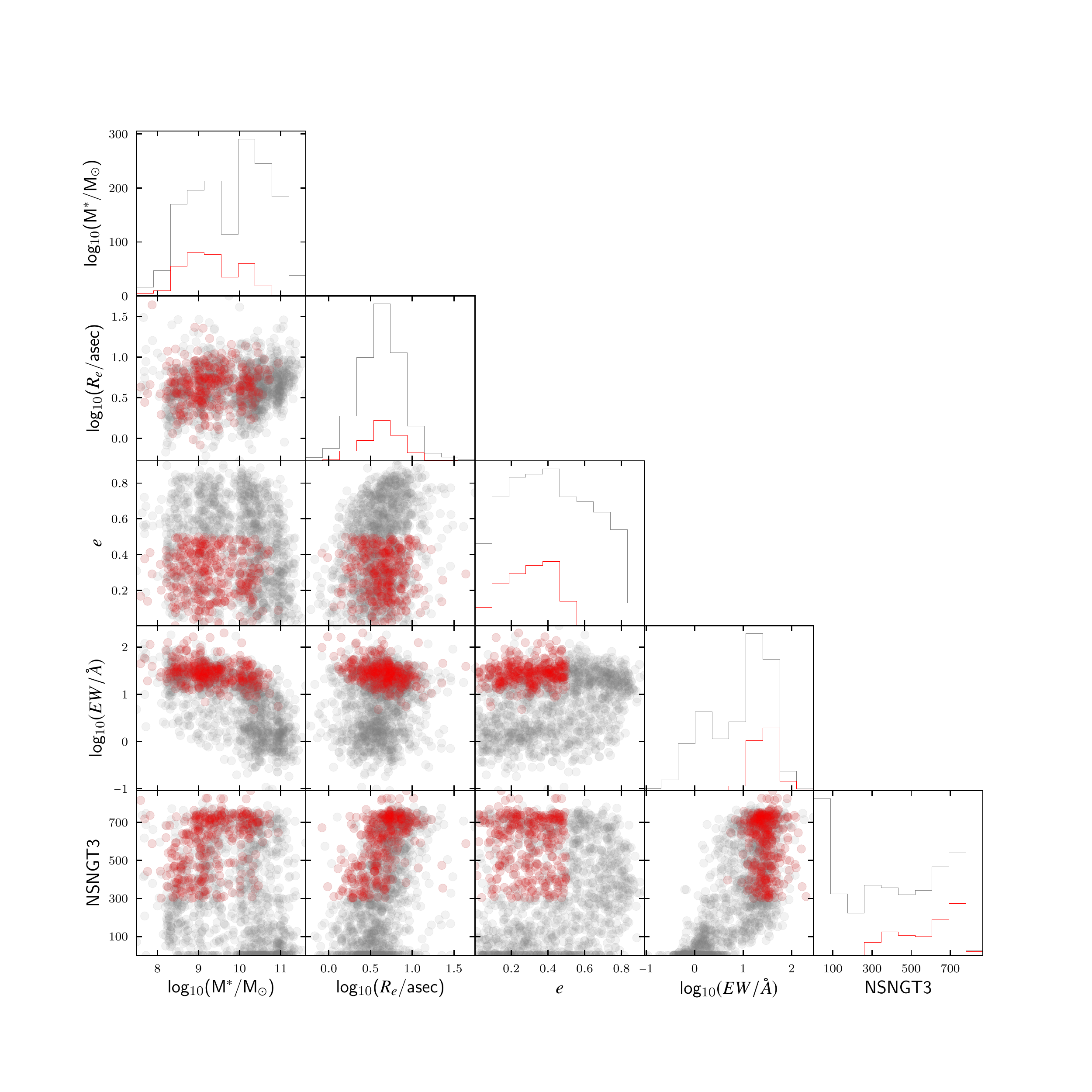} \\
    \end{center}
    \caption{Galaxy parameters for our sample of 342 galaxies (red) selected from the total SAMI Galaxy Survey (grey). We show the marginalised (diagonal) and conditional (off-diagonal) distributions for the stellar mass (log$_{10}$(M$^*$/M$_\odot$)), effective radius (log$_{10}(R_e$/asec)), ellipticity ($e = 1 - b/a$), H$\alpha$ equivalent width (log$_{10}$($EW$/\AA{})), and NSNGT3. NSNGT3 corresponds to the number of spaxels that meet our signal-to-noise masking criteria. We select a sample of star-forming galaxies from the SAMI Galaxy Survey with inclination and signal-to-noise cuts that can be adequately modelled using \blobby{}.}
    \label{fig:sample}
\end{figure*}

There are 1523 galaxies in the SAMI Galaxy Survey DR2 where all of the above diagnostic criteria are measurable. 342 galaxies remain once our criteria is applied. Figure \ref{fig:sample} shows that we are selecting galaxies along the star-forming main sequence. We see a clear bimodal distribution in the log equivalent width, where we have selected those galaxies with $EW$ > 3 \AA{}. The equivalent width cut removes massive galaxies that are typically passive, which can be seen when plotting the equivalent width compared to M$_*$ and $R_e$. There are a limited number of galaxies in our sample with 3 \AA{} < $EW$ $\lesssim$ 10 \AA{} as many of those galaxies are removed due to being classified as LINER/AGN or having < 300 spaxels that meet our signal-to-noise masking criteria.

Removing highly inclined galaxies results in a large cut to our sample, but does not bias our sample along any galaxy properties. Also, the selection of galaxies with at least 300 unmasked spaxels does remove galaxies with $R_e \lesssim 1''$, but there are very few of these galaxies in the underlying SAMI Galaxy Survey DR2 sample.

\subsection{DYNAMO sample}
\label{subsec:dynamo}

The DYnamics of Newly Assembled Massive Objects \citep[DYNAMO,][]{Green2014} survey consists of a sample of star-forming galaxies in the local Universe ($z \lesssim 0.1$). These galaxies were classified as star-forming in the MPA-JHU Value Added Catalog from the Sloan Digital Sky Survey \citep[SDSS,][]{York2000}. 
The galaxies comprising the DYNAMO survey were chosen primarily based on H$\alpha$ luminosity. The aim was to include both high H$\alpha$ luminious galaxies, that are rare in the local Universe, as well as a sample of typical galaxies in the local Universe. The resulting galaxy sample ranged \mbox{SFR $\in [1, 100]$ \smassyr{}}.

The data for the DYNAMO samples was obtained via observations using the 3.9 m Anglo-Australian Telescope (AAT) and the ANU 2.3 m Telescope at Siding Spring Observatory. The AAT was equiped with the SPIRAL Integral-Field Unit (IFU) with the AAOmega Spectrograph \citep{Sharp2006}. SPIRAL is an array of $32 \times 16$ square, 0.7$''$ lenslets with a contiguous integral field of $22.4'' \times 11.2''$. The 1700I grating was used on the red spectrograph providing a nominal resolution power of $R \sim 12000$. The ANU 2.3 m Telescope was equiped with the Wide-Field Spectrograph \citep[WiFeS,][]{Dopita2007}. WiFeS has a $25'' \times 38''$ FoV with either $1.0'' \times 0.5''$ or $1.0'' \times 1.0''$ spaxels. The I7000 grating was chosen for the red arm, which has a $6893 - 9120$ \AA{} wavelength range with a spectral resolving power of $R \sim 7000$.

A total of 67 galaxies comprised the original DYNAMO sample. We remove galaxies observed at $i > 60^\circ$, where $i$ has been measured using the SDSS photometric pipeline using an exponential disc fit to the $r$-band. We perform the same masking criteria as described for the galaxies from the SAMI Galaxy Survey. We also remove galaxies with less than 30 unmasked spaxels. 41 galaxies were retained from the original DYNAMO sample.

\section{Methods}
\label{sec:methods}

\subsection{Modelling the gas disc kinematics}
\label{subsec:method_vdisp}

We use \blobby{} \citep{Varidel2019} to infer the intrinsic gas kinematics for the observed galaxies. \blobby{} is a forward-fitting disc modelling procedure. It assumes that the gas lies in a regularly rotating thin-disc. The prior for the spatial gas distribution within the disc allows for clumpy gas profiles using a hierarchical Gaussian mixture-model. The model is constructed in 3D (position -- position -- wavelength space) and then convolved in accordance with the PSF and instrumental broadening by the LSF. The convolved model is then compared to the observed data cube.

The advantage of \blobby{} is that it is capable of performing inference for the spatial gas distribution, including substructure, plus the gas kinematics simultaneously. This is important as the effect of beam smearing is a function of the spatial gas distribution being blurred per spectral slice. As such, the observed gas kinematics is a complex function of the intrinsic spatial gas distribution, the velocity profile, and the velocity dispersion plus instrumental broadening and beam smearing. For example, \citet{Varidel2019} found that it is possible to observe spurious substructure in the gas kinematics in a symmetric regularly rotating disc with an asymmetric spatial gas distribution plus beam smearing.

Previous testing of \blobby{} has found that it is well optimised to infer the intrinsic velocity dispersion of galaxies \citep{Varidel2019}. \blobby{} was compared to an alternative forward-fitting methodology known as \bbarolo{} \citep{DiTeodoro2015}. It was also compared to other heuristic modelling approaches that have been used in the literature, such as estimating the velocity dispersion in the outskirts of the galaxy \citep[e.g.][]{Zhou2017}, correcting the observed velocity dispersion as a function of the velocity gradient \citep[e.g.][]{Varidel2016}, and subtracting the velocity gradient in quadrature from the observed velocity dispersion \citep[e.g.][]{Oliva-Altamirano2018}. \blobby{} was found to infer the intrinsic velocity dispersion more accurately than these alternative methods, particular for galaxies where the PSF or velocity gradient were greatest.

The parameterisation for \blobby{} is set within the Bayesian framework. The joint prior distribution for the parameters, hyperparameters, and data were defined in \citet{Varidel2019}. We only make minor changes to the priors that were previously proposed. We outline the motivation for changing some of the prior distributions below.

The joint prior distribution used for this work performs inferences for the H$\alpha$ flux plus the [\ion{N}{ii}]/H$\alpha$ emission flux ratio for each spatial Gaussian flux profile (often referred to as a `blob' in \blobby{}). The gas kinematics have been assumed to be consistent across the different gas components. Therefore, the inferences for the kinematics are constrained using extra information from the [\ion{N}{ii}] emission lines at 6548.1 \AA{} and 6583.1 \AA{}. The ratio of the flux between the [\ion{N}{ii}] emission lines is assumed to be $F_{6583.1} / F_{6548.1} = 3$.

To simplify the inference for the velocity dispersion, we assume a constant velocity dispersion across the disc ($\sigma_{v, 0}$). We assume no radial gradient as the results for some galaxies returned large positive gradients when using the prior suggested by \citet{Varidel2019}. The large spatial gradients in velocity dispersion after convolution appeared to be over-fitting for wider-tailed non-Gaussian emission line profiles. Therefore, we removed the velocity dispersion gradient from the inference in order to robustly infer the constant velocity dispersion component for the large sample of galaxies that were studied in this work.

We have also widened the bounds for our priors for the systemic velocity ($v_\text{sys}$) and the asymptotic velocity ($v_c$) in order to model a larger set of galaxies than was performed by \citet{Varidel2019}. Our new priors are,
\begin{align}
    v_\text{sys} & \sim \text{Cauchy}(0, 30 \text{ km s}^{-1})T(-300 \text{ km s}^{-1}, 300 \text{ km s}^{-1}), \\
    v_c & \sim \text{Loguniform}(1 \text{ km s}^{-1}, 1000 \text{ km s}^{-1}).
\end{align}
Where $T(a, b)$ represents the distribution being truncated to the interval [$a, b$].

\subsubsection{Mitigating the effects of beam smearing}
\label{subsubsec:b3d_smear}

The effect of beam smearing by the PSF is accounted for in \blobby{} by convolving the underlying model constructed by the PSF, prior to calculating the likelihood function. The PSF profile assumed in \blobby{} is a superposition of 2D concentric circular Gaussian profiles. Therefore, the PSF needs to first be modelled assuming this flux profile.

The SAMI Galaxy Survey pipeline provides estimates for the PSF by fitting a profile to a star that was observed simultaneously with the galaxy. We have used the Moffat profile estimates, where the PSF is described as,
\begin{equation}
    p(r) = \frac{\beta - 1}{\pi \alpha^2}
        \bigg(1 + \frac{r^2}{\alpha^2}\bigg)^{-\beta}.
\end{equation}
$\alpha$ is the FWHM and $\beta$ is a shape parameter that controls the tails of the Moffat profile. 

To refactor the Moffat profile parameters into a set of concentric Gaussians, we construct the 1D Moffat profile, then fit it with two 1D Gaussians. Two Gaussians were enough to adequately model the PSF profile. The estimated Gaussian parameters are then passed to \blobby{}.

For the DYNAMO sample, the FWHM of the PSF was measured during observations. As such, we assumed a 2D circular Gaussian profile to be representative of the PSF for the DYNAMO sample. Thus, the underlying model in \blobby{} was convolved with a Gaussian profile prior to comparing the model to the data for our galaxies from the DYNAMO survey.

\subsubsection{Continuum substraction}
\label{subsubsec:mask}

\blobby{} requires the data to be continuum subtracted. For galaxies from the SAMI Galaxy Survey, we use the continuum models made available in the SAMI Galaxy Survey DR2 pipeline. The full description for the continuum modelling routine is described in \citet{Owers2019}. We estimate the continuum for the galaxies from the DYNAMO survey using a 300 bin moving median filter as also implemented by \citet{Green2014}.

It is possible for the continuum modelling to introduce systematics in the resulting continuum subtracted data cube. These systematics may not be well accounted for in the \blobby{} approach. We make the assumption that the stellar continuum will be adequately modelled in regions of high H$\alpha$ signal-to-noise. This is a significant motivation for implementing the H$\alpha$ signal-to-noise masking outlined in Section \ref{subsec:Sample}.

\subsubsection{Posterior optimisation}
\label{subsubsec:posterioroptimisation}

We use \dnest{} \citep{Brewer2011DNEST,Brewer2018DNest4} to get a point estimate of the maxima for the posterior PDF. \dnest{} is a sampling algorithm based on nested sampling \citep{Skilling2004}, where the new levels are constructed by exploring a weighted mixture of the previous levels. Exploration of the levels is performed using a Metropolis Markov Chain Monte Carlo (MCMC). The multi-level exploration allows \dnest{} to be significantly more robust to local maxima compared to typical nested sampling, allowing for the exploration of high parameter spaces and multi-modal posterior distributions. Estimated values throughout this paper are of the maximum posterior PDF value in the chain sampled using \dnest{}.

\subsection{Global velocity dispersion}
\label{subsec:globalvdisp}

\subsubsection{Beam smearing corrections}
\label{subsubsec:vdispcorr}

\begin{figure}
    \begin{center}
        \includegraphics[
            width=0.5\textwidth,
            keepaspectratio=true,
            trim=17mm 9mm 0mm 16mm,
            clip=true]{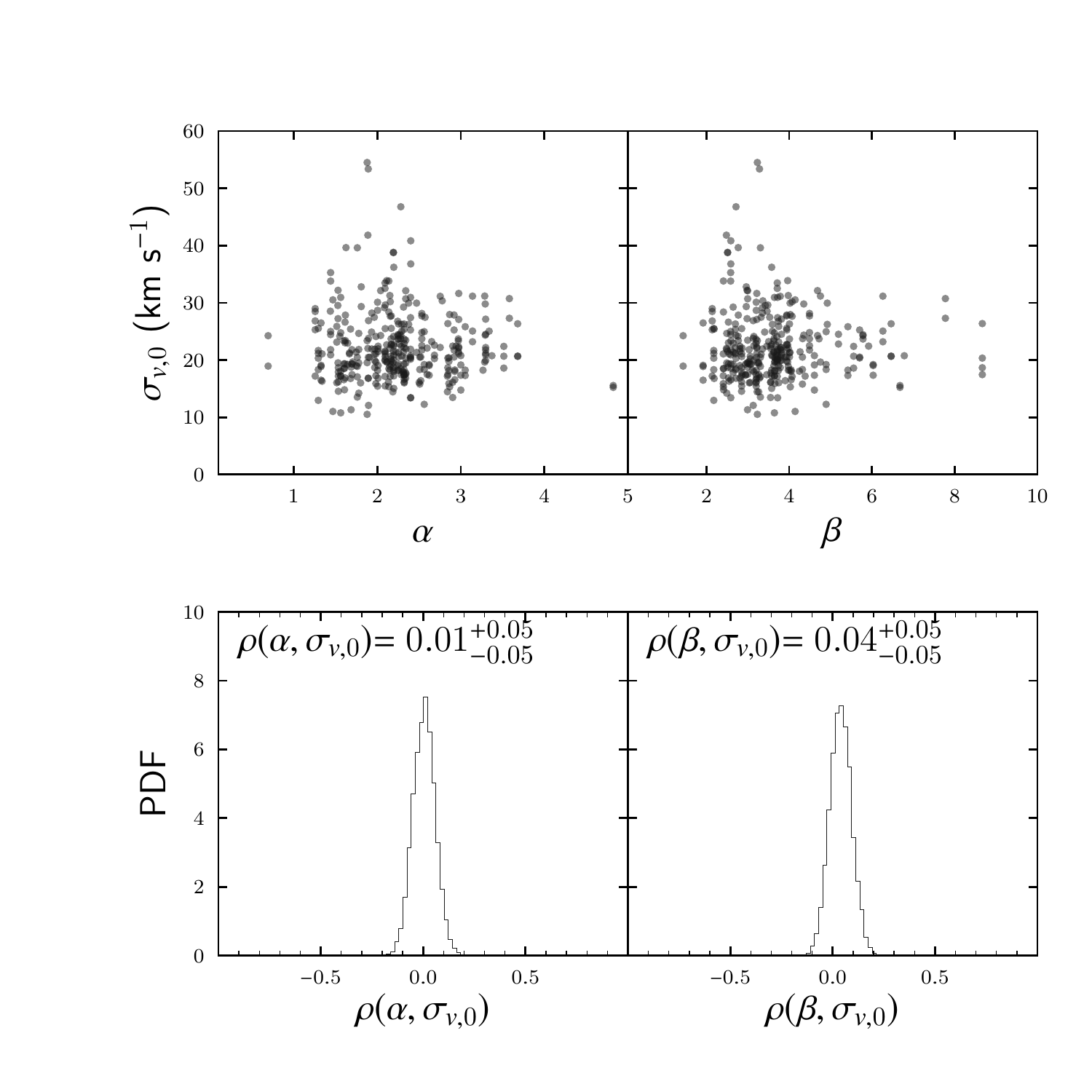} \\
    \end{center}
    \caption{
    Comparing the PSF Moffat profile parameters $\alpha$ and $\beta$ to the inferred global velocity dispersion for galaxies in our sample from the SAMI Galaxy Survey. We also show the PDF of the Spearman-rank correlation coefficients estimated using 10${^4}$ bootstrap samples (bottom). $\rho = 0$ lies within the 68\% shortest credible intervals suggesting that $\sigma_{v, 0}$ is adequately corrected for beam smearing.}
    \label{fig:psf_vdisp}
\end{figure}

Assuming that \blobby{} accurately corrects for beam smearing, there should be no residual correlation between the PSF profile parameters and the inferred intrinsic velocity dispersion ($\sigma_{v, 0}$). The distribution of $\sigma_{v, 0}$ is consistent with our expectations for a beam smearing corrected sample. Figure \ref{fig:psf_vdisp} shows a comparison between the PSF Moffat profile parameters and $\sigma_{v, 0}$ for our sample from the SAMI Galaxy Survey. For both $\alpha$ and $\beta$, zero remains inside the 68\% shortest credible intervals for the Spearman-rank correlation coefficients.

For galaxies from the DYNAMO survey, the Spearman-rank correlation coefficient is estimated as $\rho(\text{FWHM}, \sigma_v) = 0.10^{+0.17}_{-0.17}$. As zero remains within the 68\% confidence interval, this result is also consistent with a beam smearing corrected sample.

We also compare $\sigma_{v, 0}$ to an estimate of the velocity dispersion that was not corrected for beam smearing ($\sigma_{v, \text{uncorrected}}$). The uncorrected estimator is calculated as the arithmetic mean velocity dispersion across the FoV, when fitting a single Gaussian component to each spaxel. Spaxels with H$\alpha$ signal-to-noise < 3 are masked in this process to eliminate the effects of poorly constrained spaxels on the final estimate.

\begin{figure}
    \begin{center}
        \includegraphics[
            width=0.5\textwidth,
            keepaspectratio=true,
            trim=2mm 6mm 8mm 15mm,
            clip=true]{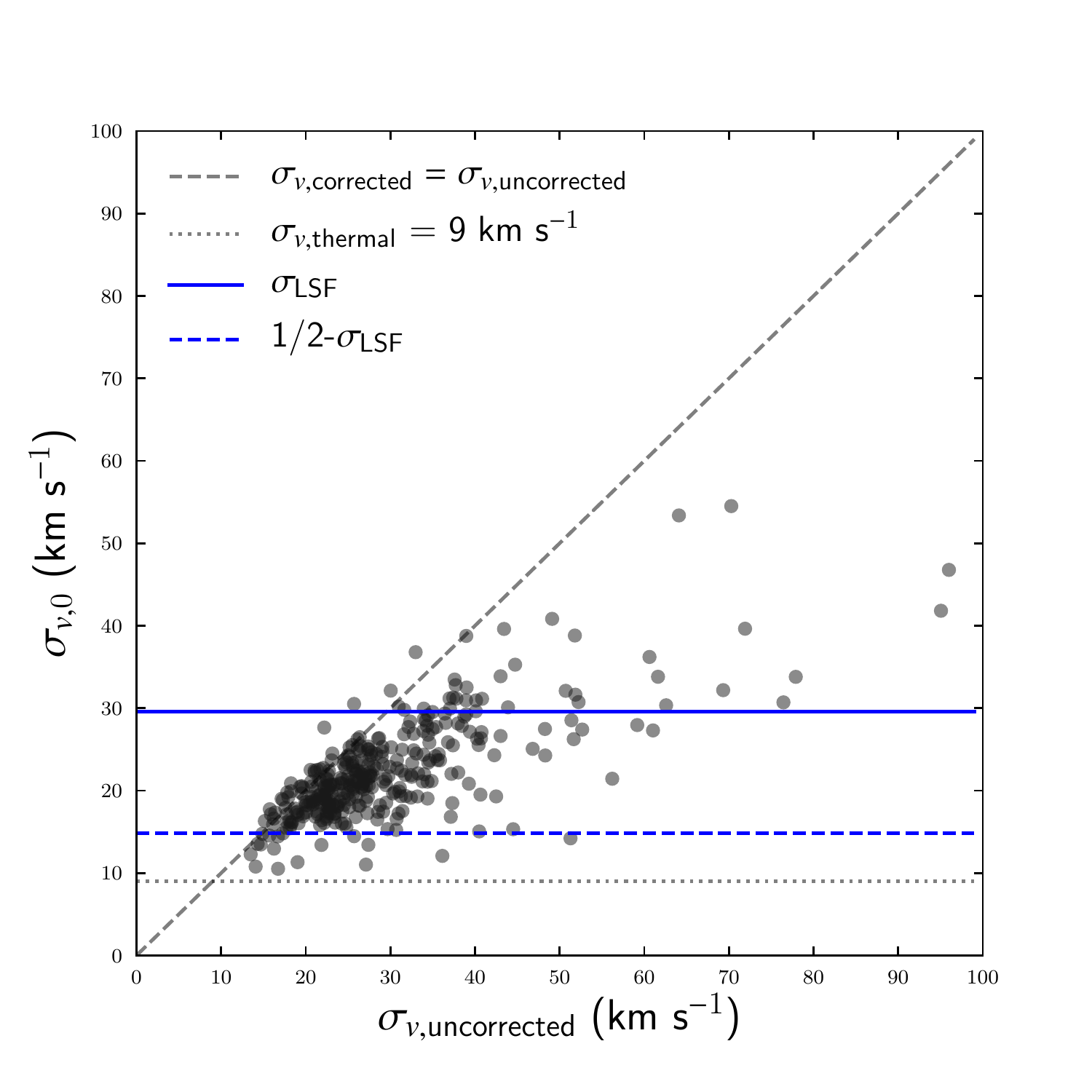} \\
    \end{center}
    \caption{$\sigma_{v, 0}$ estimated using \blobby{} compared to the arithmetic mean of the single-component fits per spaxel ($\sigma_{v, \text{uncorrected}}$) to each galaxy from the SAMI Galaxy Survey sample. Estimates for the velocity dispersion are typically lower using \blobby{} as it mitigates the effects of beam smearing.}
    \label{fig:vdisp_correction}
\end{figure}

Estimates for $\sigma_{v, 0}$ are significantly lower than $\sigma_{v, \text{uncorrected}}$ (see Figure \ref{fig:vdisp_correction}). Using the sample of galaxies from the SAMI Galaxy Survey, typical corrections were $\Delta \sigma_v = -5.3^{+4.0}_{-7.0}$ \kms{} and $\Delta \sigma_v / \sigma_{v, 0} = -0.20^{+0.14}_{-0.18}$, where $\Delta \sigma_v = \sigma_{v, 0} - \sigma_{v, \text{uncorrected}}$. The typical beam smearing corrections are consistent with the results found by \citet{Varidel2019} on a sample of 20 star-forming galaxies in the SAMI Galaxy Survey using \blobby{}. 

All estimated values have $\sigma_{v, 0} > \sigma_{v, \text{thermal}}$ = 9 \kms{}. $\sigma_{v, \text{thermal}}$ is the typical emission line width expected for a \ion{H}{ii} region at $\sim 10^4$ K \citep{Glazebrook2013}. As such, $\sigma_{v, \text{thermal}}$ sets a physically motivated lower bound.

\subsubsection{Considerations of the effects of the LSF on the velocity dispersion estimates}
\label{subsubsec:lsf}

The SAMI instrument has the spectral resolution of \mbox{\sigmalsf{} = 0.68 \AA{}} ($\sigma_{v, \text{LSF}}$ = 29.6 \kms{}) in the red arm. For reference, we show the 1-\sigmavlsf{} and 1/2-\sigmavlsf{} on Figure \ref{fig:vdisp_correction}. 89\% of our galaxies have estimated intrinsic velocity dispersions less than \sigmavlsf{} and 4.6\% of our sample were estimated to have intrinsic velocity dispersion less than \sigmavlsf{}/2.

We correct for the LSF by convolving the emission line by a Gaussian profile with $\sigma_{v, \text{LSF}}$ during the fitting procedure in \blobby{}. This procedure assumes that the observed emission line is a convolution of two Gaussians. Therefore, the estimated velocity dispersion can be affected by non-Gaussianities in the shape of the LSF, particularly when the velocity dispersion is significantly less than the width of the LSF. However, deviations of the SAMI LSF from a Gaussian profile are minor \citep{vandeSande2017}. Also 95.4\% of our sample were estimated to be $\sigma_{v, 0} >$ \sigmavlsf{}/2, as such the effects of minor systematic differences of the LSF from a Gaussian profile is unlikely to have significant effects on our inferences.

Similarly, the effect of variations in the LSF FWHM are minor for the SAMI Galaxy Survey. The LSF FWHM varied at the $\sim$5\% level as a function of fibre, time, and wavelength during the SAMI Galaxy Survey \citep{Scott2018}. For the velocity dispersions values that we estimate, this should result in uncertainties on the level of \mbox{$\Delta \sigma_{v}$ $\sim$ 1 \kms{}}. As such, the variation of the LSF FWHM is not expected to have any significant effect on the conclusions drawn in this paper.

\subsubsection{Estimating the vertical velocity dispersion}
\label{subsubsec:sigmavz}

Our disc modelling approach calculates a global estimate for the intrinsic line-of-sight (LoS) velocity dispersion ($\sigma_{v, 0} \equiv \sigma_{v, \text{LoS}}$). Most studies using IFS observations report \sigmavlos{}. However,  $\sigma_{v, \text{LoS}}$ is a mixture of the radial ($\sigma_{v, R}$), azimuthal ($\sigma_{v, \phi}$), and vertical ($\sigma_{v, z}$) velocity dispersion components.

At any point in the sky, \sigmavlos{} is given by \citep[e.g. Equation 27a, ][]{Cappellari2019},
\begin{equation}
    \sigma^2_{v, \text{LoS}} = 
        \big( \sigma^2_{v, R} \sin^2 \phi + \sigma^2_{v, \phi} \cos^2 \phi \big) \sin^2 i
        + \sigma^2_{v, z} \cos^2 i.
    \label{eq:sigmavlosdecomp}
\end{equation}
Observed \sigmavlos{} is the luminosity-weighted integral along the LoS. To calculate the average velocity dispersion, we make the following approximations. We assume that the flux is constant across a thin disc with finite radial extent. We also assume spatially constant velocity dispersion components and that $\sigma^2_{v, \perp} \equiv \sigma^2_{v, R} \approx \sigma^2_{v, \phi}$ then the average LoS velocity dispersion is given by,
\begin{equation}
    \bar\sigma^2_{v, \text{LoS}} = 
        \sigma^2_{v, \perp} \sin^2 i + \sigma^2_{v, z} \cos^2 i.
    \label{eq:sigmavlosdecompint}
\end{equation}
Setting $\gamma^2 = \sigma^2_{v, z} / \sigma^2_{v, \perp}$ , and rearranging, then
\begin{equation}
    \sigma_{v, \text{LoS}} = \sigma_{v, z} \sqrt{\sin^2 i / \gamma^2 + \cos^2 i}
    \label{eq:sigmavzsolve}
\end{equation}
The above model predicts changing \sigmavlos{} as a function of $i$ if $\gamma \neq 1$. For $\gamma > 1$, \sigmavlos{} increases with increasing $i$, whereas \sigmavlos{} decreases with $i$ when $\gamma < 1$.

To estimate $\gamma$ we assume that \sigmavlos{} follows a loguniform distribution with mean $\sigma_{v, z, 0}$ and log variance $\tau^2$. The generating function for a single data point $\sigma_{v, z, i}$ is then,
\begin{equation}
    p(\sigma_{v, \text{LoS}, j} | \sigma_{v, z, 0}, \tau^2, \gamma) \sim 
        \text{lognormal}(\sigma_{v, z, 0} \sqrt{\sin^2 i / \gamma^2 + \cos^2 i}, \tau^2).
    \label{eq:evdisp_like}
\end{equation}
We assume the following priors,
\begin{align}
    &p(\sigma_{v, z, 0}) \sim \text{loguniform}(1, 100) \\
    &p(\tau) \sim \text{loguniform}(10^{-3}, 1) \\
    &p(\gamma) \sim \text{loguniform}(0.1, 10).
    \label{eq:evdisp_prior}
\end{align}
The posterior distribution is then given by,
\begin{multline}
    p(\sigma_{v, z0}, \tau, \gamma | \mathbf{D}) =
        p(\sigma_{v, z, 0}) p(\tau) p(\gamma)
        \prod^N_{j=1} p(\sigma_{v, \text{LoS}, j} | \sigma_{v, z, 0}, \tau^2, \gamma).
    \label{eq:evdisp_posterior}
\end{multline}
The above formulation assumes independence of the prior distribution between $\sigma_{v, z, 0}$, $\tau$, $\gamma$, as well as all $\sigma_{v, \text{LoS}, j}$. The above posterior distribution can now be sampled using typical techniques. We used \textsc{emcee} to sample the posterior distribution \citep{Foreman-Mackey2013}.

\begin{figure}
    \begin{center}
        \includegraphics[
            width=0.5\textwidth,
            keepaspectratio=true,
            trim=0mm 4mm 0mm 2mm,
            clip=true]{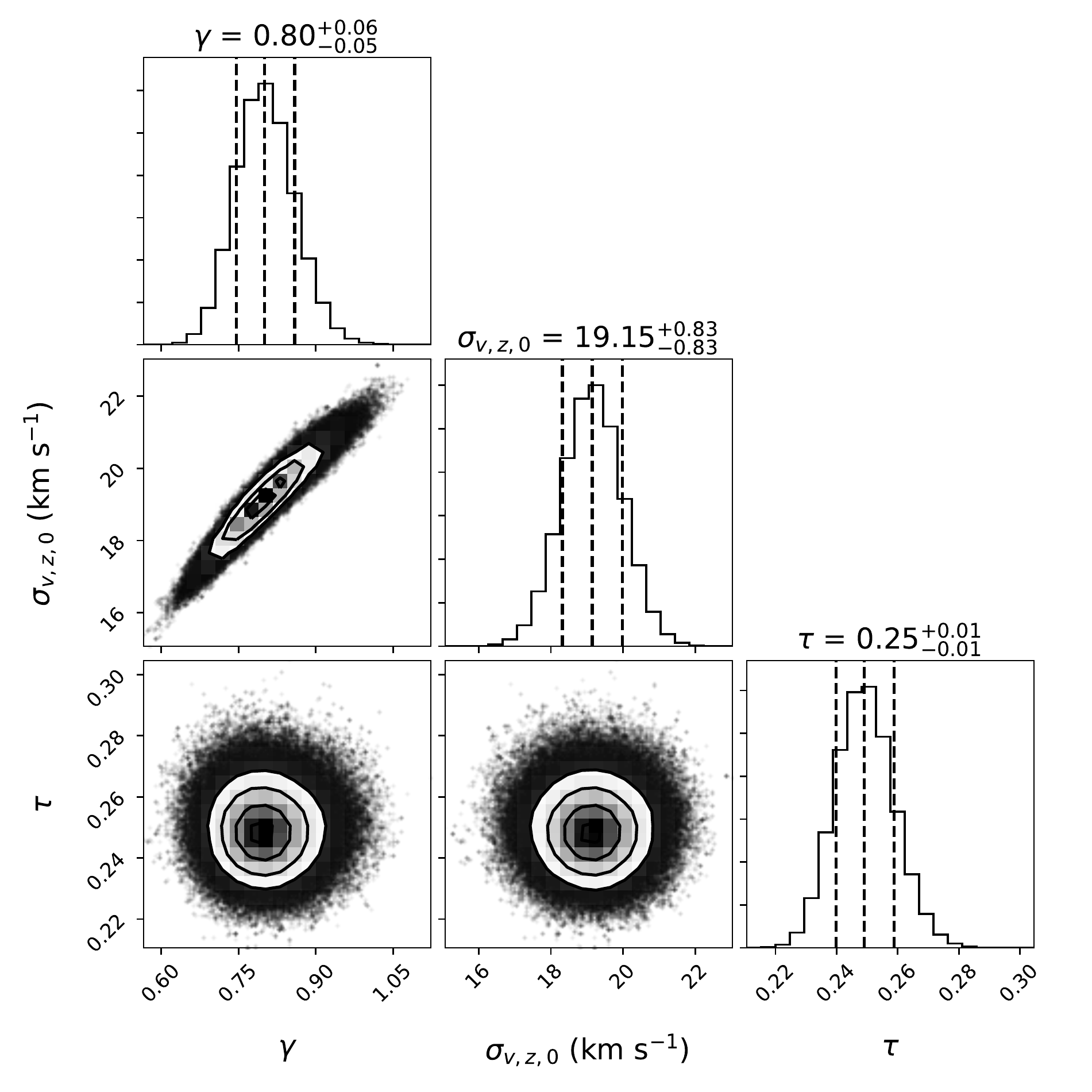} \\
    \end{center}
    \caption{Corner plot \citep{Foreman-Mackey2016} showing the marginalised (diagonal) and joint (off-diagonal) posterior distributions for the parameter estimation for the inclination dependence model. There is evidence for a dependence of \sigmavlos{} on inclination for our sample of galaxies from the SAMI Galaxy Survey. This suggests that the vertical velocity dispersion (\sigmavz{}) is less than the averaged azimuthal and radial velocity dispersion ($\sigma_{v, \perp}$).}
    \label{fig:evdisp_samples}
\end{figure}

We estimate $\gamma = 0.80^{+0.06}_{-0.05}$ as shown in Figure \ref{fig:evdisp_samples}, suggesting that the vertical velocity dispersion is less than the averaged azimuthal and radial components. This analysis was consistent with other approaches that we applied. For example, the bootstrapped Spearman-rank correlation coefficient distribution between the inclination and \sigmavlos{} was $\rho(i, \sigma_{v, \text{LoS}}) = 0.18^{+0.05}_{-0.05}$, where the uncertainties for the Sperman-rank correlation coefficient is estimated as the 68\% shortest credible interval after bootstrap resampling. We also performed the above analysis using uniform priors for $\sigma_{v, z, 0}$ and $\gamma$ with the same ranges, yet we still find $\gamma = 0.80^{+0.06}_{-0.06}$.

Previous studies have suggested that $\sigma_{v, z}/\sigma_{v, R} \sim 0.6$ \citep[Section 1.2.2, ][]{Glazebrook2013} for stars. Mean \ion{H}{i} gas velocity dispersion was reported up to $\sim 3$ times higher for galaxies observed at $i > 60^\circ$ compared to $i < 60^\circ$ by \citet{Leroy2008}, also suggesting that the contribution of $\sigma_{v, R}$ and $\sigma_{v, \phi}$ dominates.

Studies of gas kinematics have typically not reported or found evidence that $\sigma_{v, z}$ is related to the inclination. For example, studies of high-$z$ in the KMOS3D Survey have found no significant correlation between the axis ratio $q = b/a$ and \sigmavlos{} \citep{Wisnioski2015,Ubler2019}. However, such a relation may be difficult to identify in high-$z$ galaxies with lower signal-to-noise and spatial resolution.

We estimate the vertical velocity dispersion ($\sigma_{v, z}$) for individual galaxies by inverting Equation \ref{eq:sigmavzsolve} and using $\gamma$ = 0.8. We estimated the Spearman-rank correlation between the inclination and \sigmavz{} to be $\rho(i, \sigma_{v, z}) = 0.00^{+0.05}_{-0.05}$ after performing the correction per galaxy, suggesting that our analysis appropriately removed the correlation as a function of the inclination angle.

\begin{figure}
    \begin{center}
        \includegraphics[
            width=0.5\textwidth,
            keepaspectratio=true,
            trim=17mm 9mm 0mm 16mm,
            clip=true]{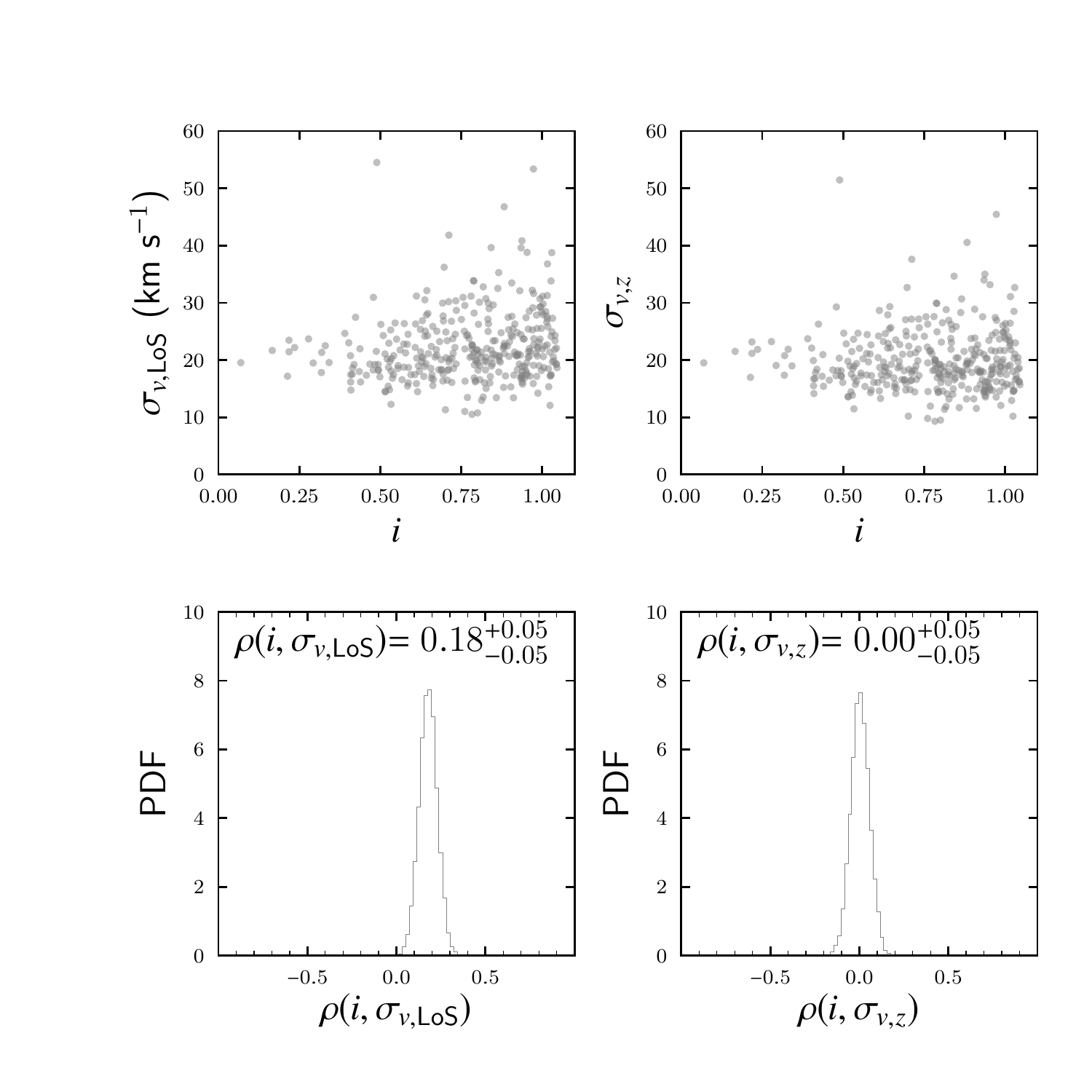} \\
    \end{center}
    \caption{The relationship between the inclination ($i$) and inferred velocity dispersion estimates. We also show the PDF of the Spearman-rank correlation coefficients using bootstrap resampling (bottom). There is evidence for a weak positive correlation between the LoS velocity dispersion \sigmavlos{} and $i$. Whereas the distribution for the vertical velocity dispersion after applying a correction factor yields no relation with $i$.}
    \label{fig:globalvdisp}
\end{figure}

\begin{figure}
    \begin{center}
        \includegraphics[
            width=0.5\textwidth,
            keepaspectratio=true,
            trim=5mm 8mm 8mm 18mm,
            clip=true]{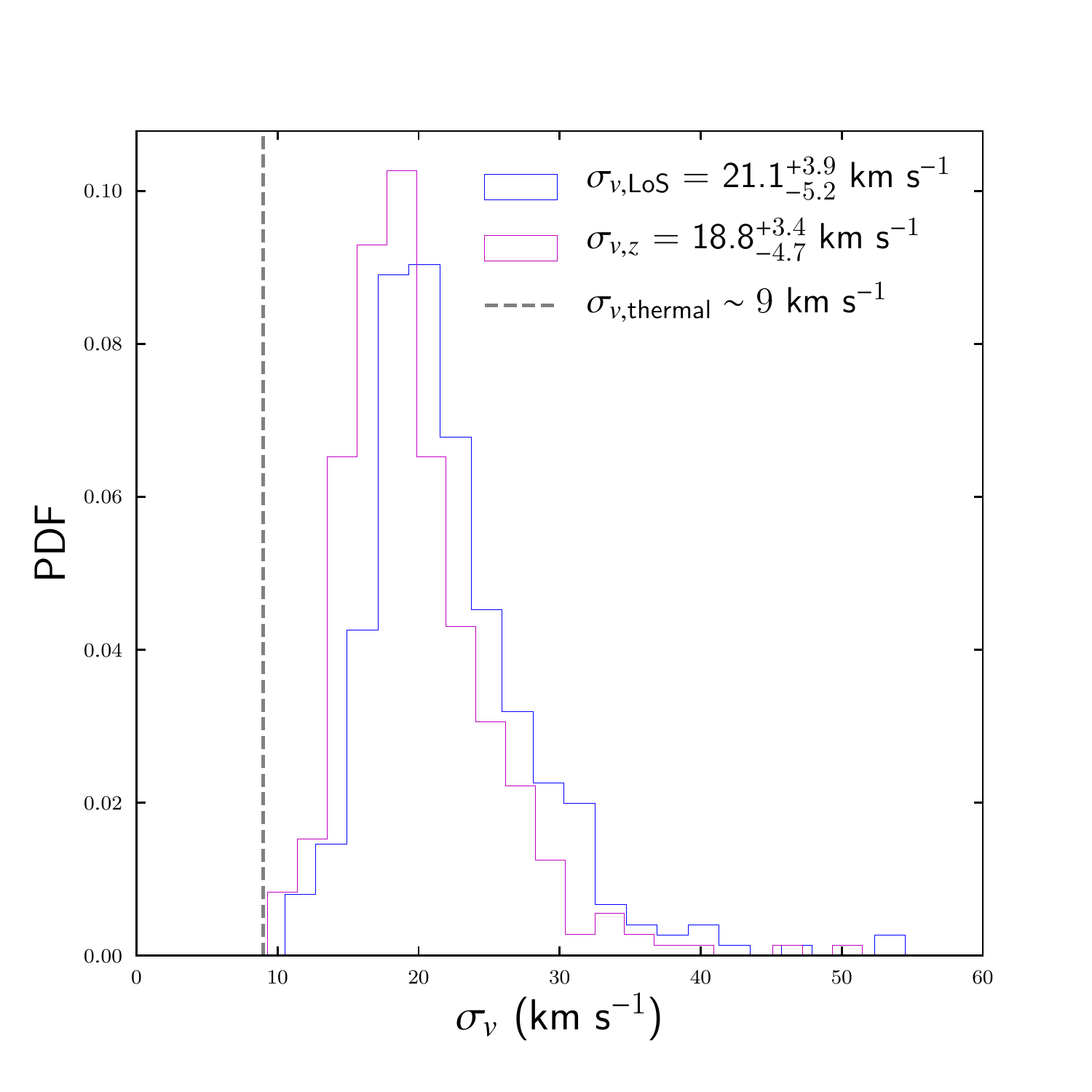} \\
    \end{center}
    \caption{The distribution of the LoS (\sigmavlos{}, blue) and vertical (\sigmavz{}, red) velocity dispersion for our sample of galaxies from the SAMI Galaxy Survey. The estimated vertical velocity dispersion is adjusted down with respect to \sigmavlos{} by a couple of \kms{} in accordance with the inclination correction described in Section \ref{subsubsec:sigmavz}.}
    \label{fig:vdisp_hist}
\end{figure}

Converting from \sigmavlos{} to \sigmavz{} adjusts the reported values by a couple of \kms{}. The marginalised distributions yield \mbox{\sigmavlos{} = 21.1$^{+3.9}_{-5.2}$ \kms{}} and \mbox{\sigmavz{} = 18.8$^{+3.4}_{-4.8}$ \kms{}} (see Figure \ref{fig:vdisp_hist}). Typical differences are \mbox{$\sigma_{v, \text{LoS}} - \sigma_{v, z} = 2.4^{+0.9}_{_-1.3}$ \kms{}}, with the greatest correction being \mbox{$\sigma_{v, \text{LoS}} - \sigma_{v, z} = 7.9$ \kms{}}.

For the remainder of this paper, we will report the values of \sigmavz{}. The subsequent analysis and results do not change qualitatively whether we use \sigmavz{} or \sigmavlos{}, but \sigmavz{} is preferred as it is an estimator free from effects from the viewing angle. It is also more appropriate to compare \sigmavz{} to theoretical models, as they are typically framed with respect to \sigmavz{}. We report both values in Appendix \ref{appendix:parameters}.

We have not applied the inclination correction for galaxies observed in the DYNAMO survey. This is due to finding no significant relation with $\rho(i, \sigma_{v, \text{los}}) = -0.09^{+0.15}_{-0.15}$ for our galaxies from the DYNAMO survey. This suggests that there is no inclination effect to correct for within this sample. It may be that the sample from the DYNAMO survey is too small to infer the inclination effect. In this case, we choose not to apply the inclination effect found from the SAMI Galaxy Survey, as it is still possible that the inferred effect is methodological rather than physical across all galaxies.

\subsection{Circular velocity estimates}
\label{subsec:vcirc}

\blobby{} estimates the LoS velocity profile using the empirical model proposed by \citet{Courteau1997},
\begin{equation}
    v(r) = 
        v_\text{c}
        \frac{(1 + r_t/r)^\beta}{(1 + (r_t/r)^\gamma)^{1/\gamma}}
        \sin(i) \cos(\theta)
        + v_{\text{sys}}.
    \label{eq:vprof}
\end{equation}
Where $v_c$ is the asymptotic velocity and $v_\text{sys}$ is the systemic velocity. $r$ is defined by the distance to the kinematic centre. $r_t$ is the turnover radius. $\beta$ is a shape parameter that controls the gradient for $r > r_t$, where the velocity gradient increases for $\beta < 0$, and decreases when $\beta > 0$. $\gamma$ is a shape parameter that controls how sharply the velocity profile turns over. $i$ is the inclination of the galaxy. Then $\theta$ is the polar angle in the plane of the disc.

We intend to estimate the circular velocity from our inferred parameters. While $v_c$ is a natural choice, it is difficult to get a strong constraint on $v_c$ across our complete sample due to the FoV for the SAMI Galaxy Survey typically extending out to $\sim$1.5 $R_e$. Instead, we estimate the absolute circular velocity at 2.2 $R_e$ denoted as $v_{2.2}$ following \citep{Bloom2017a}.

For low values of $i$, small differences in the estimated $i$ can result in large difference of $v_{2.2}$. Therefore, for low values of $i$, incorrect estimates for the observed ellipticity can result in large changes in our estimates for the inclination. As such, we restrict our calculated values for $v_{2.2}$ to galaxies in the range $i \in [30^\circ, 60^\circ]$ ($e \in [0.13, 0.5]$ assuming a thin disc). 

Similarly, galaxies with $R_e < 3.0''$ tended to have very large scatter on their $v_{2.2}$. At these limits, the spatial resolution of our observations are likely playing a role in increasing the scatter in the rotational velocity estimates. 230 galaxies meet the above inclination and $R_e$ criteria. We only reference $v_{2.2}$ for galaxies that meet that inclination for the remainder of this paper. 

\subsection{Integrated star-formation rates}
\label{subsec:sfr}

We used the best fit SFRs from the GAMA Survey \citep{Gunawardhana2013,Davies2016,Driver2018}. The SFRs are estimated using full spectral energy distribution (SED) fitting of 21 bands of photometry across the UV, optical, and far infrared ranges with \textsc{MAGPHYS} \citep{daCunha2008}. \textsc{MAGPHYS} fits the observed photometry using a library that includes stellar spectral and dust emission profiles. In this way, the SFRs are corrected for dust emission. These estimates for the SFR were used instead of the SAMI H$\alpha$ luminosity maps as there are known aperture affects given the limited FoV of the SAMI instrument \citep[Appendix A,][]{Medling2018}.

For the galaxies from the DYNAMO survey, we used the SFR values reported by \citet{Green2014}. SFRs were estimated using the H$\alpha$ luminosity estimated from their observations. The SFR estimate includes a dust correction using the Balmer decrement from the ratio between their measured H$\alpha$ and H$\beta$ measurements. The SFR was then calculated using the dust-corrected H$\alpha$ luminosity maps that were converted to SFR maps using the \citet{Kennicutt1994} conversion assuming a \citet{Chabrier2003} IMF.

\subsection{Integrated \ion{H}{i} gas measurements}
\label{subsec:totalhi}

Follow-up 21 cm observations of SAMI galaxies were obtained as part of the SAMI-HI survey, carried out with the Arecibo
radio telescope (Catinella et al. in prep.). Observations and data reduction were analogous to those of the xGASS survey \citep{Catinella2018}, with the only difference that these were not gas fraction-limited observations. We observed each galaxy until detected, but moved to another target if there was no hint of \ion{H}{i} signal within the first 20-25 minutes of on-source integration. 

\ion{H}{i} emission-line spectra were obtained for 153 galaxies with these dedicated follow-up observations; on-source integration times ranged between 2 and 50 minutes, with an average of 15 minutes. Together with an additional 143 good HI detections (i.e., classified as detection code `1') in the Arecibo Legacy Fast ALFA \citep[ALFALFA][]{Giovanelli2005,Haynes2018} survey, SAMI-\ion{H}{i} includes global \ion{H}{i} spectra for 296 SAMI galaxies from the SAMI Galaxy Survey catalogue. 95 galaxies overlap with our sample selection from the SAMI Galaxy Survey.

\section{Results}
\label{sec:results}

\subsection{Low gas velocity dispersion in the SAMI Galaxy Survey}
\label{subsec:vdisp_sami}

We find vertical velocity dispersions lower than previously reported for studies of the gas kinematics in the SAMI Galaxy Survey. The median vertical velocity dispersion is \mbox{\sigmavz{} = 18.8 \kms{}} for our sample as shown in Figure \ref{fig:vdisp_hist}. The 68-th shortest credible interval is \mbox{[14.1, 22.1] \kms{}} and the 95-th shortest credible interval is \mbox{[11.4, 30.0] \kms{}}. The maximum inferred vertical velocity dispersion for a single galaxy is \sigmavz{} = 51 \kms{}. We now compare this to two other studies of the gas kinematics of galaxies from the SAMI Galaxy Survey by \citet{Zhou2017} and \citet{Johnson2018}.

Analysing 8 star-forming galaxies in the SAMI Galaxy Survey, \citet{Zhou2017} found that 7 out of 8 galaxies had \mbox{$\sigma_\text{gas} \in [20, 31]$ \kms{}}. Their remaining galaxy (GAMA 508421) was reported as \mbox{$\sigma_\text{gas} = 87 \pm 44$ \kms{}}. GAMA 508421 exhibits a high circular velocity in the outskirts of the SAMI FoV \mbox{($v \sim 130$ \kms{})} and a clear centralised peak in velocity dispersion that is typical of beam smearing affected galaxies. Our estimate for GAMA 508421 is \mbox{\sigmavz{} = 22 \kms{}}. As such, we suspect that the reported velocity dispersion for GAMA 508421 is greater than it's intrinsic velocity dispersion.

The discrepancy between \citet{Zhou2017} and our estimates, particularly with GAMA 508421, is most likely due to the different beam smearing corrections. \citet{Zhou2017} report the flux-weighted mean velocity dispersion using spaxels where \mbox{$\sigma_v > 2 v_\text{grad}$}. $v_\text{grad}$ is an estimate for the local velocity gradient using adjacent spaxels defined as \citep{Varidel2016},
\begin{equation}
    v_\text{grad}(x, y) = \sqrt{(v(x + 1) - v(x-1))^2 + (v(y + 1) - v(y - 1))^2}.
    \label{eq:vgrad}
\end{equation}
See Section 5.1.1 by \citet{Varidel2019} for a revised calculation of the velocity gradient using a finite-difference scheme.

The approach used by \citet{Zhou2017} usually removes the centre of the galaxies, where the velocity gradient is steepest. This approach results in a significant downward correction compared to the uncorrected velocity dispersion estimates. However, the outskirts of galaxies can still be affected by beam smearing. Also, it is possible that the centre of the galaxy may be effected by beam smearing, yet not reach the \mbox{$\sigma_v > 2 v_\text{grad}$} criteria, which is likely to have occurred in the case of GAMA 508421. The approach of \citet{Zhou2017} was also shown previously to over-estimate the intrinsic velocity dispersion in toy models \citep[Section 5.1.1.,][]{Varidel2019}

Another study of a sample of 274 star-forming galaxies from the SAMI Galaxy Survey was performed by \citet{Johnson2018}. They removed galaxies with \mbox{$M_* > 8 \times 10^{10}$ M$_\odot$} and S\'ersic index of $n > 2$. They also removed galaxies that they deem to be spatially unresolved or have kinematic uncertainties greater than 30\%. While they do not provide summary statistics for their inferred velocity dispersion values from the SAMI Galaxy Survey, their plots show a typical range of \mbox{$\sigma_0 \in [20, 60]$ \kms{}}, plus one galaxy at \mbox{$\sigma_0 \sim 90$ \kms{}}. This is slightly above our range of velocity disperisions.

\begin{figure*}
    \begin{center}
        \includegraphics[
            width=\textwidth,
            keepaspectratio=true,
            trim=15mm 15mm 15mm 15mm,
            clip=true]{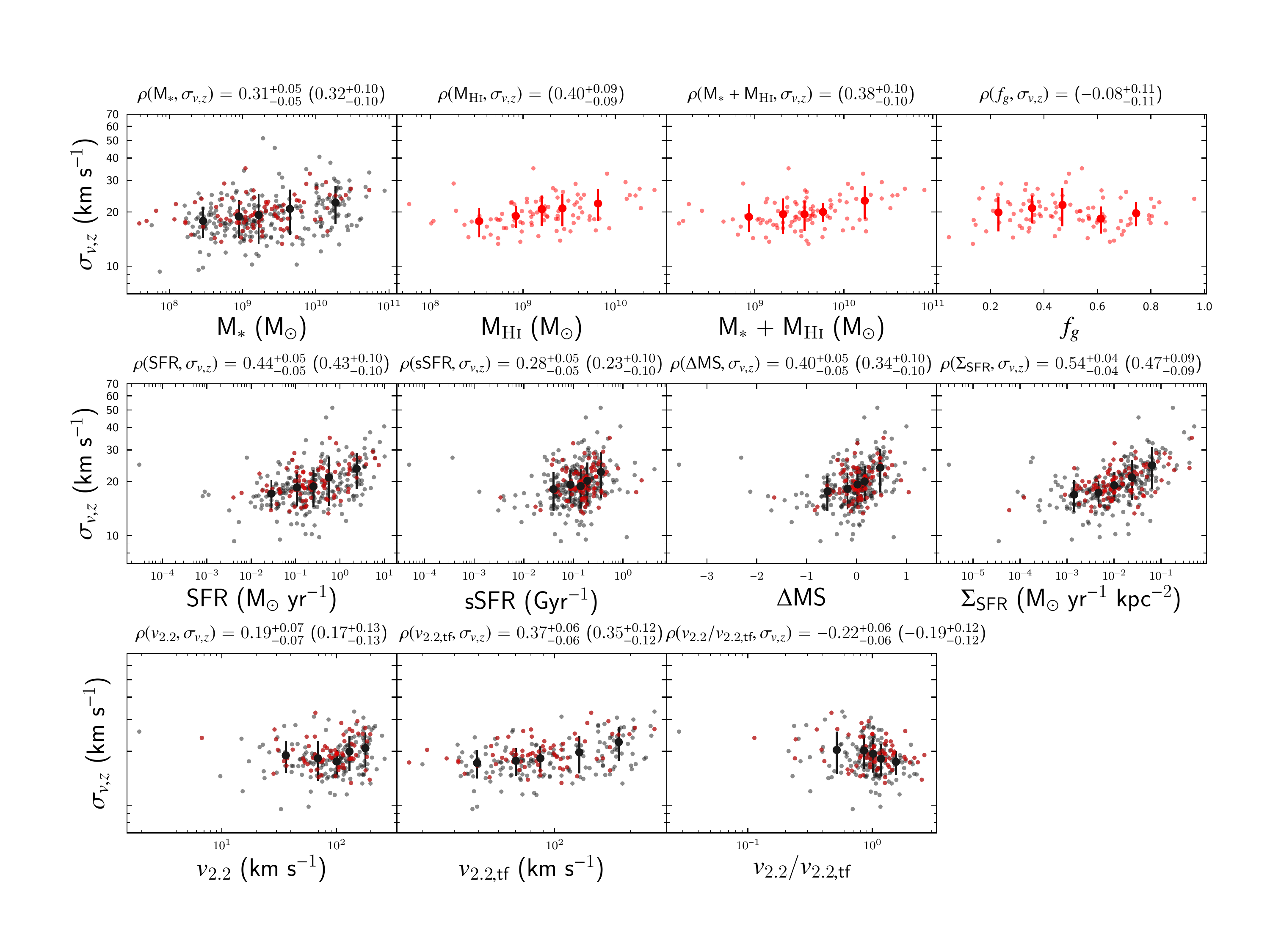} \\
    \end{center}
    \caption{Comparing global intrinsic vertical velocity dispersion (\sigmavz{}) to global properties for galaxies from the SAMI Galaxy Survey. We show the relation of \sigmavz{} with measures of mass (top), star-formation rate (middle), and rotational velocity (bottom), respectively. Red points indicate the galaxies with observed integrated \ion{H}{i} masses. The Spearman-rank correlation coefficients are shown at the top of each plot, with brackets indicating the correlation coefficient for galaxies with measured \ion{H}{i} masses. The uncertainties for the Spearman-rank correlation coefficients are estimated as the 68\% shortest credible interval from 10$^4$ bootstrapped samples. We find significant positive correlations with measures of mass, star-formation rate, and rotational velocity. The greatest positive correlation we find is with star-formation rate surface density (\sfrdensity{}).}
    \label{fig:globaltrends}
\end{figure*}

To estimate the intrinsic velocity dispersion, \citet{Johnson2018} calculated the median velocity dispersion across the kinematic maps or at the outskirts of their galaxy. They then apply a further correction on their estimated velocity dispersion by using a lookup table of toy galaxies that have been constructed with beam smearing effects. The slight difference between our studies may be driven solely by their choice of using a single FWHM estimate for the PSF rather than the Moffat profile used in this paper. Also, increased scatter may occur in their estimator due to being affected by low signal-to-noise spaxels in the outskirts of the galaxies.

\subsection{Correlation of global velocity dispersion and integrated star-formation rate}
\label{subsec:vdisp_sfr}

Correlation analysis between the global velocity dispersion and several global galaxy properties from the SAMI Galaxy Survey reveals that \sigmavz{} has the greatest positive correlation with star-formation rate measures (Figure \ref{fig:globaltrends}). We estimate the Spearman-rank correlation between the SFR and \sigmavz{} to be $\rho$(SFR, \sigmavz{}) = $0.44^{+0.05}_{-0.05}$. We control for several factors in order to investigate this relationship further.

The correlation between \sigmavz{} and star-formation rate increases when accounting for the galaxy size. To do this, we estimate the average star-formation rate surface density, \sfrdensity{} = SFR/$\pi{} R_e^2$ where $R_e$ is the effective radius. The Spearman-rank correlation is then $\rho$(\sfrdensity{}, \sigmavz{}) = $0.54^{+0.04}_{-0.04}$. Velocity dispersion is expected to increase with star-formation rate surface density assuming that star-formation feedback processes are acting as a driver of turbulence \citep[e.g.][]{Ostriker2011,Faucher-Giguere2013}. As such, this does provide support that star-formation feedback processes is acting as a driver of turbulence within this sample of galaxies.

Figure \ref{fig:globaltrends} also shows a positive correlation between \sigmavz{} and integrated stellar mass (M$_*$), \ion{H}{i} gas mass (M$_\text{\ion{H}{i}}$), as well as the sum of M$_*$ and M$_\text{\ion{H}{i}}$. Interestingly, there is a suggestion that M$_\text{\ion{H}{i}}$ is slightly more correlated than M$_*$ with \sigmavz{}, although the uncertainties are wide enough that we cannot confirm that is the case. SFR is well known to be correlated with M$_*$, which adds a further complication in determining the relation between \sigmavz{} and SFR.

To account for the SFR -- M$_*$ relation, we calculated the specific star-formation rate (sSFR = SFR/M$_*$) and $\Delta$MS. $\Delta$MS is calculated as the log difference between the SFR and the star-forming main sequence relation as proposed by \citet{Renzini2015}. We find that the correlation between \sigmavz{} and star-formation rate decreased after accounting for stellar mass. This suggests that the relation between \sigmavz{} and star-formation rate is a combination of both SFR and stellar mass related quantities.

Despite the correlation between \sigmavz{} and star-formation rate estimators, the absolute change in \sigmavz{} as a function of SFR remains slight across the dynamic range of SFR $\in [10^{-3}, 10]$ \smassyr{}. We report the change in velocity dispersion in 5 SFR bins in Table \ref{tab:surveys_sfr_vdisp}. The change in mean velocity dispersion between the end bins from SFR = 0.029 \smassyr{} to SFR = 2.4 \smassyr{} is only 6.41 \kms{}. A similarly shallow gradient was found by \citet{Johnson2018} using data from the SAMI Galaxy Survey.

\begin{table*}
    \caption{Comparing summary statistics of the vertical velocity dispersion in other samples compared to those in this work. Each sample was split into 5 bins of equal percentile widths. We show the mean ($\bar\sigma_{v, z}$), standard deviation ($\Delta$\sigmavz{}), the standard error ($\Delta \bar\sigma_{v,z}$), median (med(\sigmavz{})), and bootstrap resampled standard deviation of the median ($\Delta$med(\sigmavz{})). The groups of galaxies are as follows: Low-$z$ (H$\alpha$) \citep{Epinat2008,Moiseev2015}, \ion{H}{i} surveys where 15 \kms{} has been added in-quadrature \citep{Leroy2008,Walter2008,Ianjamasimanana2012,Stilp2013}, high-$z$ analogues from \citet{Varidel2016} plus the re-analysed galaxies from the DYNAMO survey, plus high-$z$ (H$\alpha$) \citep{Johnson2018,Cresci2009,Wisnioski2011,Epinat2009,Jones2010,DiTeodoro2016}.
    }
    \begin{center}
    \begin{tabular}{lccccccc}
    \hline
Group & Bin & SFR (\smassyr{}) & $\bar\sigma_{v, z}$ (\kms{}) & $\Delta$\sigmavz{} (\kms{}) & $\Delta \bar\sigma_{v,z}$ (\kms{}) & med(\sigmavz{}) (\kms{}) & $\Delta$med(\sigmavz{}) (\kms{}) \\
    \hline
SAMI (H$\alpha$)
    & 1 & 0.029 & 17.12 & 3.21 & 0.39 & 17.13 & 0.29 \\
    & 2 & 0.11 & 18.54 & 3.99 & 0.49 & 18.31 & 0.41 \\
    & 3 & 0.25 & 18.79 & 4.34 & 0.53 & 18.52 & 0.43 \\
    & 4 & 0.57 & 21.07 & 6.47 & 0.79 & 19.72 & 0.71 \\
    & 5 & 2.4 & 23.54 & 5.35 & 0.65 & 23.54 & 0.64 \\
    \hline
Low-$z$ (H$\alpha$)
    & 1 & 0.0047 & 19.46 & 2.89 & 0.43 & 18.84 & 0.72 \\
    & 2 & 0.046 & 20.77 & 4.33 & 0.65 & 19.21 & 0.41 \\
    & 3 & 0.18 & 20.57 & 3.86 & 0.58 & 19.21 & 0.6 \\
    & 4 & 0.37 & 21.66 & 4.55 & 0.68 & 19.85 & 0.44 \\
    & 5 & 1.0 & 23.5 & 7.0 & 1.0 & 21.21 & 0.81 \\
    \hline
Low-$z$ (H\textsc{i})
    & 1 & 0.0014 & 16.95 & 0.55 & 0.18 & 16.86 & 0.15 \\
    & 2 & 0.005 & 17.39 & 0.64 & 0.20 & 17.44 & 0.25 \\
    & 3 & 0.066 & 18.65 & 2.98 & 0.99 & 17.81 & 0.6 \\
    & 4 & 0.58 & 19.18 & 1.36 & 0.43 & 18.78 & 0.57 \\
    & 5 & 2.2 & 20.82 & 2.58 & 0.82 & 19.9 & 1.4 \\
    \hline
High-$z$
    & 1 & 0.96 & 27.0 & 3.2 & 1.1 & 26.23 & 0.94 \\
Analogues (H$\alpha$) 
    & 2 & 3.2 & 39.4 & 12.6 & 4.4 & 40.0 & 4.9 \\
    & 3 & 9.1 & 40.7 & 14.3 & 5.0 & 41.2 & 7.8 \\
    & 4 & 17 & 43.0 & 15.2 & 5.4 & 42.9 & 7.6 \\
    & 5 & 27 & 55.9 & 15.6 & 5.2 & 54.8 & 5.4 \\
    \hline
High-$z$ (H$\alpha$)
    & 1 & 3.4 & 44.0 & 20.5 & 1.6 & 39.8 & 1.9 \\
    & 2 & 6.4 & 45.8 & 18.2 & 1.5 & 43.1 & 1.2 \\
    & 3 & 10 & 44.3 & 20.3 & 1.6 & 42.8 & 3.2 \\
    & 4 & 20 & 48.3 & 20.2 & 1.6 & 45.0 & 1.5 \\
    & 5 & 82 & 53.2 & 20.0 & 1.6 & 51.0 & 2.6 \\
    \hline
    \end{tabular}
    \label{tab:surveys_sfr_vdisp}
    \end{center}
\end{table*}

Galaxies are often kinematically classified as either rotationally or turbulence dominated by comparing the ratio of rotational and random velocities ($v/\sigma$). In a similar vain to such analysis, we also investigated the relation between \sigmavz{} and rotational velocity. \sigmavz{} is shown compared to the rotational velocity measures using \blobby{} ($v_{2.2}$) as outlined in Section \ref{subsec:vcirc} and using the Tully-Fisher relation \citep[$v_{2.2, \text{tf}}$,][]{Bloom2017b}. 

We find a positive correlation between \sigmavz{} and the rotational velocity estimators. This is to be expected as rotational velocity is also correlated with stellar mass. To control for that effect, we calculated the ratio between $v_{2.2}$ and $v_{2.2, \text{tf}}$. We then find a negative correlation between \sigmavz{} and $v_{2.2} / v_{2.2, \text{tf}}$. As such, we observe that galaxies exhibit greater rotation than their mass predicts when \sigmavz{} is lesser, and lesser rotation when \sigmavz{} is greater.

\subsection{Comparisons with other surveys}
\label{subsec:surveys}

\begin{figure*}
    \begin{center}
        \includegraphics[
            width=\textwidth,
            keepaspectratio=true,
            trim=3mm 8mm 5mm 10mm,
            clip=true]{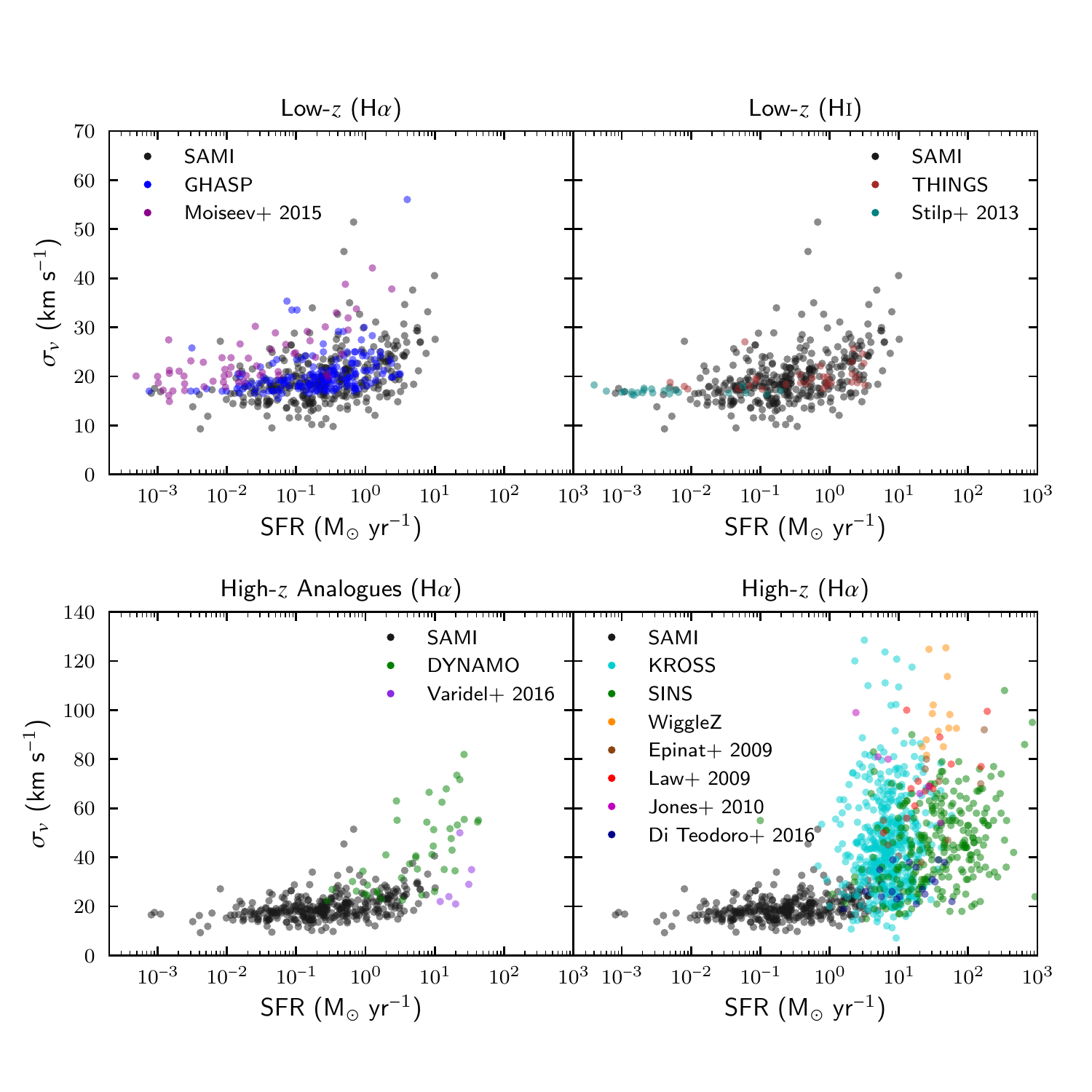} \\
    \end{center}
    \caption{Comparison of the SFR -- velocity dispersion (\sigmav{}) relation compared to others surveys in the literature. The sets of galaxies that constitute each subplot are the same as outlined in Table \ref{tab:surveys_sfr_vdisp}. We find the SFR -- \sigmav{} relation increases slightly across the range SFR $\in [10^{-3}, 1]$ \smassyr{}, then turns up significantly at SFR $\gtrsim$ 1 \smassyr{}. This relation is approximately consistent across all surveys.
    }
    \label{fig:sfr_vdisp_compar}
\end{figure*}

In this section we aim to describe our results from the SAMI Galaxy Survey in the context of other studies. In Table \ref{tab:surveys_sfr_vdisp} and Figure \ref{fig:sfr_vdisp_compar} we show comparisons of velocity dispersion compared to SFR. The data is shown in four groups of galaxies; low-$z$ measured using H$\alpha$ \citep{Epinat2008,Moiseev2015}, low-$z$ measured using \ion{H}{i} \citep{Leroy2008,Walter2008,Ianjamasimanana2012,Stilp2013}, High-$z$ analogues from \citet{Varidel2016} plus the galaxies that we re-analysed from the DYNAMO sample, and high-$z$ galaxies at z $\gtrsim$ 1 \citep{Johnson2018,Cresci2009,Wisnioski2011,Epinat2009,Law2009,Jones2010,DiTeodoro2016}. Table \ref{tab:sami_compare} also outlines qualitative ranges for the galaxy parameters for galaxies at low-$z$ measured using the H$\alpha$ emission line, including other studies of the SAMI and DYNAMO samples.

The comparative data sets have been measured using both ionised and neutral gas. For ionised gas, there are two additional contributions to the velocity dispersion. One is the thermal broadening of \mbox{$\sigma_\text{thermal} \sim 9$ \kms{}}, corresponding to the typical temperature of an \ion{H}{ii} region. There is also a contribution from the expansion speed of the \ion{H}{ii} region. Studies of the expansions speed reveal \mbox{$\sigma_\text{expand} \sim 10$ \kms{}} for small regions, up to \mbox{$\sigma_\text{expand} \sim 13 - 17$ \kms{}} for larger regions \citep{Chu1994}.

Given the contributions of $\sigma_\text{thermal}$ and $\sigma_\text{expand}$ to the observed ionised gas kinematics, we perform several adjustments to the comparative velocity dispersion estimates. For ionised gas estimates, we remove any corrections for the additional contributions. For \ion{H}{i} studies, we assume a nominal contribution due to these effects of 15 \kms{}, that we add in quadrature to the published velocity dispersion estimates. We note that in other studies, 15 \kms{} has been subtracted in quadrature from the ionised gas measurements for comparisons between different studies. We prefer the alternative as 15\% of our galaxies have \mbox{\sigmavz{} < 15 \kms{}}.

\subsubsection{Comparison with surveys at low-$z$}
\label{subsubsec:lowz_surveys}

The SAMI Galaxy Survey has similar selection criteria to the Mapping Nearby Galaxies at Apache Point Observatory \citep[MaNGA,][]{Bundy2015} survey in terms of fundamental galaxy properties (see Table \ref{tab:sami_compare}). Our data have similar ranges in redshifts, stellar mass, and SFR. As such, we would naively expect the gas turbulence within our sample to be similar to the MaNGA survey estimates.

\begin{table*}
    \caption{Qualitative ranges of galaxy parameters for low-$z$ samples in the literature, where gas kinematics were estimated using the H$\alpha$ emission line.}
    \begin{center}
        \begin{tabular}{lcccc}
            \hline
            Sample & $z$ & log$_{10}$(M$_*$/M$_\odot$) & log$_{10}$(SFR / \smassyr{}) & $\sigma_v$ (\kms{}) \\
            \hline
            SAMI (this work) & [0.005, 0.08] & [7.5, 11] & [-3, 1] & [10, 60] \\
            SAMI \citep{Johnson2018} & $< 0.1$ & [7.5, 11] & [-3, 1] & [20, 90] \\
            SAMI \citep{Zhou2017} & $< 0.1$ & [9.8, 10.8] & - & [20, 90] \\
            DYNAMO (this work) & [0.06, 0.15] & [9, 11] & [-1, 2] & [10, 80] \\
            DYNAMO \citep{Green2014} & [0.06, 0.15] & [9, 11] & [-1, 2] & [10, 90] \\
            GHASP \citep{Epinat2008} & $\sim 0.01$ & - & [-3, 1] & [15, 30] \\
            \citet{Moiseev2015} & < 90 Mpc & - & [-3, 1] & [15, 40] \\
            \citet{Varidel2016} & [0.01, 0.04] & [10.5, 11] & [1, 1.6] & [20, 50] \\
            MaNGA \citep{Yu2019} & [0.01, 0.15] & [8.5, 11.5] & [-2, 1] & [10, 130] \\
            \hline
        \end{tabular}
        \label{tab:sami_compare}
    \end{center}
\end{table*}

We find systematically lower velocity dispersions than those estimated by \citet{Yu2019}. They estimated mean velocity dispersions of \mbox{$\sigma \in [20, 50]$ \kms{}} across various galaxy property ranges \citep[Figure 6,][]{Yu2019}. Specifically for SFR vs. velocity dispersion they found mean \mbox{$\sigma \in [30, 50]$ \kms{}} across 4 bins in the range \mbox{SFR $\in [10^{-2}, 10]$ \smassyr{}}. Whereas we estimate mean $\bar\sigma_{v, z} \in [17, 24]$ \kms{} across 5 bins of \mbox{SFR $\in [10^{-3}, 10]$}.

\citet{Yu2019} also reported galaxies with velocity dispersion of \mbox{$\sigma_v \gtrsim 50$ \kms{}} up to \mbox{$\sigma_v \sim 130$ \kms{}}. This is similar to $\sigma_v$ estimates for galaxies at high redshift (see high-$z$ galaxies, Table \ref{tab:surveys_sfr_vdisp}). However, we see very little evidence for a significant fraction of galaxies with \mbox{$\sigma_v \gtrsim 50$ \kms{}}.

The spectral resolution of \mbox{\sigmalsf{} $\in [50, 80]$ \kms{}} \citep{Bundy2015,Yu2019} may be an issue for MaNGA. The variability in the MaNGA spectral resolution could correspond to a large scatter in their estimated velocity dispersion, that may explain their upper limit of $\sigma_v \sim 100$ \kms{}. We also show that the velocity dispersion is significantly less than their spectral resolution, thus their assumptions regarding the LSF will be important.

Instead, our results are closer to the velocity dispersion estimates found in the Gassendi HAlpha survey of SPirals \citep[GHASP,][]{Epinat2008}, where their galaxies overlap in SFR. We can see in \mbox{Figure \ref{fig:sfr_vdisp_compar}} that our samples match well with the work of \citet{Epinat2008} both in terms of mean velocity dispersion and gradient as a function of SFR. We only disagree slightly in terms of the intrinsic scatter, which could be sample selection, methodology, or signal-to-noise dependent.

We highlight that \citet{Epinat2008} estimated their velocity dispersion using the residuals in spatially resolved mean velocity compared to a rotational velocity model. As such, their measurements are fundamentally different and should not be affected by $\sigma_\text{thermal}$ and $\sigma_\text{expand}$. So we added \mbox{15 \kms{}} in quadrature to their published velocity dispersion estimates for comparison purposes.

Our results are also qualitatively similar to those published by \citet{Moiseev2015}, who studied a sample of nearby dwarf galaxies. Their results agree with the higher end of our velocity dispersion estimates, although there is still an offset in the mean velocity dispersion. We note that \citet{Moiseev2015} do not explicitly correct for beam smearing, but due to studying nearby galaxies at < 90 Mpc, the effects of beam smearing should be minimal.

Combining the results of \citet{Moiseev2015} and \citet{Epinat2008}, we find differences of the mean and median velocity dispersion estimates compared to our sample of \mbox{$\sim 1-3$ \kms{}} (see Table \ref{tab:surveys_sfr_vdisp}), where our results were systematically lower. The difference of $\sim 2$ \kms{} could be explained due to calculating \sigmavz{} rather than \sigmavlos{}, which resulted in a downward shift in our velocity dispersion estimates by \mbox{$\sim 2$ \kms{}} as described in Section \ref{subsubsec:sigmavz}.

We find little difference in the intrinsic scatter between our sample and the combined samples of \citet{Moiseev2015} and \citet{Epinat2008}. Calculating the 1-sigma standard deviation for the sample ($\Delta$\sigmavz{}), sample mean ($\Delta\bar\sigma_{v, z}$), and median ($\Delta$med(\sigmavz{})), we find that all variance estimates were of similar magnitude (see Table \ref{tab:surveys_sfr_vdisp}). As such, we conclude that our results are approximately consistent with the analyses of \citet{Moiseev2015} and \citet{Epinat2008} at low-$z$ using ionised gas, albeit with different selection and methodologies in inferring the intrinsic velocity dispersion. The only exception in inferred velocity dispersions at low-$z$ using the ionised gas is the results of \citet{Yu2019} using MaNGA data where we estimate systematically lower $\sigma_v$.

Comparisons to the \ion{H}{i} observations suggest that we get the same approximately flat SFR -- \sigmav{} relation across the range SFR $\in [10^{-3}, 10]$ \smassyr{}. While there are only slight differences between the mean velocity dispersion of \mbox{$\sim 1-4$ \kms{}} across varying SFR ranges, it is important to reiterate that the \ion{H}{i} results have \mbox{15 \kms{}} added in quadrature, which is the typical difference between \ion{H}{i} and H$\alpha$ estimates for the velocity dispersion. The varying contributions of $\sigma_\text{thermal}$ and $\sigma_\text{expand}$ may cause a larger scatter than the neutral hydrogen estimates.

\subsubsection{Comparisons with surveys at high-$z$ and high-$z$ analogues}
\label{subsubsec:highz_surveys}

We now compare our results to those at high-$z$ and high-$z$ analogues. The data sets included are from the DYNAMO survey, which we have re-analysed using \blobby{}. We also include the beam-smearing corrected estimates denoted as $\sigma_{\text{m}, \text{uni}, v_g=0}$ from \citet[][]{Varidel2016}. These samples are of galaxies at low-$z$ with SFR $\gtrsim$ 1 \smassyr{}, that are similar to galaxies at high-$z$ (see Table \ref{tab:surveys_sfr_vdisp}). As such, high-$z$ analogues are likely to have similar properties to our galaxy sample at similar SFR.

Our re-analysis of the galaxies from the DYNAMO survey find results consistent with \citet{Green2014}. The difference between our results and those of \citet{Green2014} are \mbox{$\sigma_{v, z} - \sigma_{v, \text{green}} = 0.0^{+4.9}_{-6.5}$ \kms{}}. Follow-up studies of galaxies from the DYNAMO survey have also found similar results including re-analysis using alternative beam smearing corrections \citep{Bekiaris2016} and observations using adaptive optics \citep[]{Oliva-Altamirano2018}.

There is a slight increase in $\sigma_v$ when comparing SAMI with the high-$z$ analogues at overlapping SFR. At \mbox{SFR $\sim 3$ \smassyr{}}, we estimate \mbox{$\bar\sigma_{v, \text{SAMI}} = 23.54\pm0.65$ \kms{}} compared to \mbox{$\bar\sigma_{v, \text{HzA}} = 27.0\pm1.1$ \kms{}} at \mbox{SFR $\sim 2.4$ \smassyr{}} and \mbox{$\bar\sigma_{v, \text{HzA}} = 39.4\pm4.4$ \kms{}} at \mbox{SFR $\sim 3.2$ \smassyr{}} for the high-$z$ analogues. The highest velocity dispersions are primarily from the DYNAMO survey. We note that while \blobby{} was applied to both samples, the PSF for DYNAMO was assumed to be a Gaussian profile compared to a Moffat profile for the SAMI Galaxy Survey. This may result in an increased beam smearing correction in the SAMI Galaxy Survey compared to the DYNAMO survey. Also, the inclination correction was only applied to SAMI, which resulted in a $\sim$ 2 \kms{} subtraction to the initially inferred velocity dispersion from \blobby{}. As such, a difference of $\sim$ 10 \kms{} may not be significant given limitations of comparing the two samples.

The high-$z$ analogues extend the trend of increasing \sigmav{} with SFR (see Figure \ref{fig:sfr_vdisp_compar}). This trend starts to increases within the sample from SAMI Galaxy Survey at SFR $\gtrsim 1$ \smassyr{}. Expanding the star-formation rate range up to SFR $\sim$ 100 \smassyr{} using the high-$z$ analogues, we see that trend increases dramatically with $\sigma_v$ up to 80 \kms{} in the range  SFR $\in$ [10, 100] \smassyr{}, which is qualitatively consistent with samples at high-$z$.

The high-$z$ galaxies exhibit a wide range of \mbox{$\sigma_v \in [10, 150]$ \kms{}}. Some of this extent is likely to be driven by lower signal-to-noise at higher redshift. Furthermore, systematic biases such as beam smearing effects, that act to increase $\sigma_v$, will be greater due to the lower spatial resolution. Instead, the high-$z$ galaxies still exhibit similar $\sigma_v$ as the high-$z$ analogues when studied as a group.

The high-$z$ galaxies still exhibit a trend of increasing velocity dispersion as function of SFR. There is a change from \mbox{$\sigma_v$ $\sim$ 40 \kms{}} to \mbox{$\sim$ 50 \kms{}} for \mbox{SFR of 3 to 82 \smassyr{}} (see Table \ref{tab:surveys_sfr_vdisp}). We estimated the correlation to be \mbox{$\rho(\text{SFR}, \sigma_v) = 0.17^{+0.03}_{-0.04}$}. This is a weaker correlation between SFR and $\sigma_v$ than observed in low-$z$ galaxies. Lesser correlation is likely linked to increased scatter for observations of galaxies at high-$z$. The increase in scatter may be driven by signal-to-noise, beam smearing effects due to lower spatial resolution, or a change in the physical drivers of gas turbulence at high-$z$. 

There is evidence for increased $\sigma_v$ at high-$z$ compared to the high-$z$ analogues at similar SFRs. In Table \ref{tab:surveys_sfr_vdisp}, we show binned estimators for dynamic ranges of SFR $\in [3, 30]$ \smassyr{} for these two samples. $\sigma_v$ is $\sim$5 \kms{} higher at similar SFRs for the high-$z$ galaxies compared to the high-$z$ analogues.

\section{The drivers of turbulence within low-$z$ galaxies}
\label{sec:vdisp_drivers}

Turbulence in the Interstellar Medium (ISM) is expected to dissipate on the order of the disc crossing time \citep{MacLow1998,Stone1998}. Thus, an ongoing energy source is required to maintain supersonic gas turbulence across epochs. Two proposed drivers are star-formation feedback process and gravity driven turbulence.

\subsection{Star formation feedback driven turbulence}
\label{subsec:sffeedback}

Star-formation feedback processes inject momentum into the ISM through several mechanisms. These mechanisms include supernova, stellar winds, expanding \ion{H}{ii} regions, and radiation pressure from highly dense star clusters. Therefore, there has been a claim that star-formation feedback processes could provide an ongoing source of energy for the supersonic turbulence in the ISM.

Observational studies have routinely found that there is a positive correlation between global $\sigma_v$ and SFR, that has been used as evidence to support star-formation feedback processes as a driver of turbulence \citep{Green2010,Green2014,Moiseev2015,Johnson2018,Ubler2019,Yu2019}. In Section \ref{subsec:vdisp_sfr} we showed that this correlation exists in our sample of galaxies. We also showed that this correlation extends to higher SFR when connecting our sample to other galaxy surveys. 

The relationship between SFR and $\sigma_v$ has also been considered in theoretical and computational studies. Typically, the energy contribution from supernovae is considered to dominate, and therefore, has been the primary focus of most of these studies. The momentum injection per mass of stars is often assumed to be on the order of \mbox{\pmstar{} = 3000 \kms{}}. Incorporating this momentum injection into theoretical models results in assuming that the rate of momentum injection is proportional to the star-formation rate surface density, thus \mbox{$\dot{P}$ $\propto$ \pmstar{} $\Sigma_\text{SFR}$} \citep[e.g.][]{Ostriker2011,Faucher-Giguere2013,Krumholz2018}. Therefore, we expect the velocity dispersion to be positively correlated with star-formation rate surface density, if star-formation feedback processes is playing a role in driving turbulence in the ISM.

We showed in Section \ref{subsec:vdisp_sfr} that \sigmavz{} has a strong positive correlation with the galaxy averaged star-formation rate surface density. This is consistent with other analyses of the star-formation rate density and velocity dispersion \citep[e.g.][]{Lehnert2009,Yu2019,Ubler2019}. In some cases, this has been used as evidence for star-formation feedback processes acting as a primary driver of turbulence \citep{Lehnert2009,Lehnert2013}. Yet if star-formation feedback processes are acting as a driver of turbulence, we should expect that the localised \sfrdensity{} and \sigmav{} are correlated, yet some analyses have found this relation \citep{Lehnert2009,Lehnert2013}, and other studies have found a weak or statistically insignificant relation between these localised properties \citep{Genzel2011,Varidel2016,Zhou2017,Ubler2019}. Another approach to compare the observed velocity dispersion to the star-formation rate is to construct a bottom-up approach whereby \sfrdensity{} is modeled on the local scale and then integrated across the disc to estimate SFR.

To estimate \sfrdensity{} as a function of galaxy properties, it is first noted that the star-formation rate surface density is a function of the star-forming molecular gas fraction ($f_\text{sf}$) of the gas surface density ($\Sigma_\text{gas}$), that is then converted to stars at a star-formation rate efficiency per free-fall time (\eff{}). Following \citet{Krumholz2018} this can be written as,
\begin{equation}
    \Sigma_\text{SFR} = \frac{\epsilon_\text{ff}}{t_\text{ff}} f_\text{sf} \Sigma_\text{gas},
\end{equation}
where the remaining undefined quantity is the free-fall time ($t_\text{ff}$). This can then be incorporated into models to make predictions for the velocity dispersion.

One approach is to assume that the star-formation law is retained on the subgalactic scale. This assumes that \eff{} is approximately constant across the galaxy, which is broadly in agreement with the literature  \citep{Krumholz2007,Krumholz2012,Federrath2013,Salim2015,Krumholz2019}. While some studies have found evidence for varying \eff{} as a function of galaxy properties \citep{Hirota2018,Utomo2018}, the results and implications for the value of \eff{} remains in dispute. Furthemore, studies using the above approximation have found that \mbox{$\sigma_{v, z} \lesssim 25$ \kms{}}, with little variation of \sigmavz{} as a function of star-formation rate \citep{Ostriker2011,Krumholz2018}. As noted in the above samples, there is a large population of galaxies with \mbox{\sigmavz{} $\gtrsim$ 25 \kms{}}, particularly at high redshifts. As such, it is unlikely that this model is able to explain the full range of observed \sigmavz{}. Furthermore, such models allow for the variation of the Toomre $Q$ stability parameter, which leads to disagreements with observations. Hereafter, we will use the `No Transport, Fixed \eff{}' model constructed by \citet{Krumholz2018} as representative of such models.

Another approach is to assume that \eff{} can vary as a function of galaxy properties. One such approach was developed by \citet{Faucher-Giguere2013}, which assumes that the Toomre stability criteria $Q$ self-regulates to 1. In their model, when $Q < 1$ the rate of constructing giant molecular clouds (GMCs) increases, thus increasing star-formation efficiency, driving $Q$ upwards to 1. When $Q > 1$ the rate of GMC construction is limited and thus star-formation slows, leading to $Q$ decreasing to 1. The \citet{Faucher-Giguere2013} predicts that \eff{} increases with molecular gas content of the galaxy, leading to a correlation between SFR and velocity dispersion, thus potentially providing an explanation for the SFR -- \sigmav{} relation. Hereafter, we will refer to this model as `No Transport, Fixed $Q$' and use the analytical model proposed by \citet{Krumholz2018} for comparison in the following sections.

\subsection{Gravity driven turbulence}
\label{subsec:gravturbulence}

An alternative to star-formation feedback processes is driving of turbulence due to gravitational mechanisms. In such models, the gravitational potential energy of the gas is converted to kinetic energy, thus driving the turbulence in the ISM. Several mechanisms for this to occur are via accretion onto the disc, accretion through the disc, gravitational instabilities in the disc, or gravitational interactions between components of the disc. 

During the initial formation of the disc, there is evidence that accretion onto the disc can cause the high levels of gas turbulence. However, this can only be sustained on the order of the accretion time \citep[][]{Aumer2010,Elmegreen2010}. After initial disc formation, the effect of accretion onto the disc is unlikely to have a significant contribution on the gas turbulence \citep{Hopkins2013}. 

Instead, it has been shown that the supersonic turbulence initially set in the ISM during galaxy formation will quickly approach a steady-state solution \citep{Krumholz2010}. Such a steady state solution can be found where the sole driving force is due to the accretion of gas through the disc balanced by the loss of turbulence primarily by shocks. This yields prescriptions for radial models of the gas surface density and $\sigma_{v, z}$. Making simplifying assumptions whereby the entire ISM is assumed to be a single star-forming region, and integrating the models over the radial extent of the disc, they derive a relationship that simplifies to SFR $\propto$ \sigmavz{}, assuming other disc parameters are constant.

The above model is an instantaneous steady state solution, that is a function of the gas accretion rate and energy loss at the time. As the gas accretion rate has decreased over epochs, this model predicts lower gas turbulence in the ISM of galaxies at low-$z$. In Section \ref{subsubsec:highz_surveys} we highlighted that velocity dispersions were $\sim$ 5 \kms{} higher in the high-$z$ sample compared to the high-$z$ analogues sample at similar SFR. This is consistent with the velocity dispersion decreasing as a function of decreasing gas accretion rate over time. Numerous other studies have also found that gas turbulence increases as a function of $z$ \citep[][]{Kassin2012,Wisnioski2015,Johnson2018,Ubler2019}.

\subsection{Combining star-formation feedback and gravity driven turbulence}
\label{subsec:sff+grav}

\citet{Krumholz2018} recently pointed out that star-formation feedback processes can be added as an extra source of energy to the transport equation derived in \citet{Krumholz2010}. Similar to the previously mentioned models for star-formation feedback processes, they only assume the contribution of supernovae on the gas turbulence. 

Their full `Transport + Feedback' model gives a SFR -- \sigmavz{} relation of the form,
\begin{multline}
    \text{SFR} = 
        \frac{2}{1 + \beta} 
        \frac{\phi_a f_\text{sf}}{\pi G Q}
        f_{g, Q} v_c^2 \sigma_{v, z} \\
        \times 
        \max \bigg[ 
            \sqrt{ \frac{2 (1 + \beta)}{3 f_\text{g,P}}}
            \phi_\text{mp} \frac{8 \epsilon_\text{ff} f_{g, Q}}{Q}, 
            \frac{t_{\text{orb}, \text{out}}}{t_\text{sf, max}}
        \bigg]
    \label{eq:krumholz2018eq60}
\end{multline}
$f_\text{sf}$ is the fraction of the gas in the molecular star-forming phase. $f_{g, P}$ is the fractional contribution of the gas to the self-gravitation pressure at the mid-plane.  $f_{g, Q}$ is the fractional gas contribution to the toomre-$Q$ parameter. $\beta$ describes the slope of the rotation curve ($\beta = d \ln v_c / d \ln r$). $t_\text{sf, max}$ corresponds to the maximum star-formation timescale. $t_\text{orb, out}$ corresponds to the orbital period at the edge of the star-forming dominated disc. $\phi_a$ is a constant that accounts for an offset due to observing global rather than local properties, with $\phi_a = 1$ for local galaxies. $\phi_\text{mp}= 1.4$ corresponds to the assumed ratio of total pressure compared to turbulent pressure at the mid-plane.

This model results in a SFR -- \sigmavz{} relation with a floor at \mbox{15 \kms{} $\lesssim$ \sigmavz{} $\lesssim$ 25 \kms{}} (including the expansion and thermal contributions) for the lower SFR region, thus reproducing gas turbulence that is consistent with the `No Transport, Fixed \eff{}' model. The SFR -- \sigmavz{} relation then transitions to SFR $\propto$ \sigmavz{} for higher SFR, consistent with the `No Feedback' model. 

Another important contribution of \citet{Krumholz2018} is that after deriving the transport equation, they can use it to find the steady state solutions making various assumptions. The above model assumes that there is a contribution of star-formation driven turbulence ($\sigma_{v, \text{sf}}$) to the total turbulence (\sigmavz{}), where
\begin{multline}
    \sigma_{v, \text{sf}} = 
        \frac
            {4 f_\text{sf} \epsilon_\text{ff} \langle p_*/ m_* \rangle}
            {\sqrt{3 f_{g, P}} \pi \eta\phi_\text{mp} \phi_Q \phi_\text{nt}^{3/2}} \\
       \times \max \bigg[
            1, 
            \sqrt{\frac{3 f_{g, P}}{8(1 + \beta)}} 
            \frac{Q_\text{min} \phi_\text{mp}}{ 4 f_{g, Q} \epsilon_\text{ff}} \frac{t_\text{orb}}{t_\text{sf, max}}
            \bigg].
    \label{eq:sigmasf}
\end{multline}
Here $\eta = 1.5$ is a scaling parameter for the dissipation rate. $\phi_\text{mp} = 1.4$ is the ratio of total to turbulent pressure at the midplane. $\phi_Q = 2$ is the gas to stellar $Q$ plus one. By setting $\sigma_{v, \text{sf}} = 0$, \citet{Krumholz2018} derive the `No Feedback' model. In that case, the disc must remain stable, such that $Q = 1$.

\citet{Krumholz2018} derive the `No Transport, Fixed \eff{}' model by setting $\sigma_{v, z} = \sigma_{v, \text{sf}}$. In that case, the contribution is purely driven by the balance between gravitational collapse and star-formation driven by supernovae outwards. The model is similar to the model proposed by \citet{Ostriker2011}.

The `No Transport, Fixed $Q$' model, is derived by revisiting their transport equation and looking for solutions where $Q$ is set as a constant. They derive a slightly different relation given by,
\begin{equation}
    \text{SFR} = \frac{4 \eta \sqrt{ \phi_\text{mp} \phi_\text{nt}^3 } \phi_Q}{G Q^2}
        \bigg( \frac{p_*}{m_*} \bigg)^{-1}
        \frac{f_{g, Q}^2}{f_{g, P}}
        v_c^2 \sigma_{v, z}^2.
    \label{eq:ntfq}
\end{equation}
The formulation of different drivers using the same theoretical backing allows for a relatively easy comparison between the observations and different model assumptions.

\subsection{Comparison with theoretical model tracks}
\label{subsubsec:krumholztracks}

We now compare our observations to the theoretical models described above. We compare our data to the \citet{Krumholz2018} theoretical model tracks for various galaxy groups; low-$z$ dwarfs, low-$z$ spirals, and high-$z$ galaxies. For each galaxy group we use the set of parameters suggested by \citet{Krumholz2018}, which are shown in Table \ref{tab:krumholz_params}. To account for the thermal and expansion contributions to the velocity dispersion of the \ion{H}{ii} regions, 15 \kms{} was added in quadrature to the theoretical models.

\begin{table}
    \caption{Parameter values for \citet{Krumholz2018} theoretical model tracks used for Figure \ref{fig:vdispallkrumholz2018}.}
    \begin{center}
    \begin{tabular}{lcccc}
    \hline
        Parameter & Local dwarf & Local Spiral & High-$z$ \\
    \hline
        $f_\text{sf}$ & 0.2 & 0.5& 1.0 \\
        $v_c$ (\kms{}) & 100 & 220 & 200 \\
        $t_\text{orb}$ (Myr) & 100 & 200 & 200 \\
        $\beta$ & 0.5 & 0.0 & 0.0 \\
        $f_{g, Q} = f_{g, P}$ & 0.9 & 0.5 & 0.7 \\
        $\phi_a$ & 1 & 1 & 3 \\
        SFR$_\text{min}$ (\smassyr{}) & - & - & 1 \\
        SFR$_\text{max}$ (\smassyr{}) & 0.5 & 50 & - \\    
    \hline
    \end{tabular}
    \label{tab:krumholz_params}
    \end{center}
\end{table}

\begin{figure*}
    \begin{center}
        \includegraphics[
            width=\textwidth,
            keepaspectratio=true,
            trim=5mm 13mm 5mm 10mm,
            clip=true]{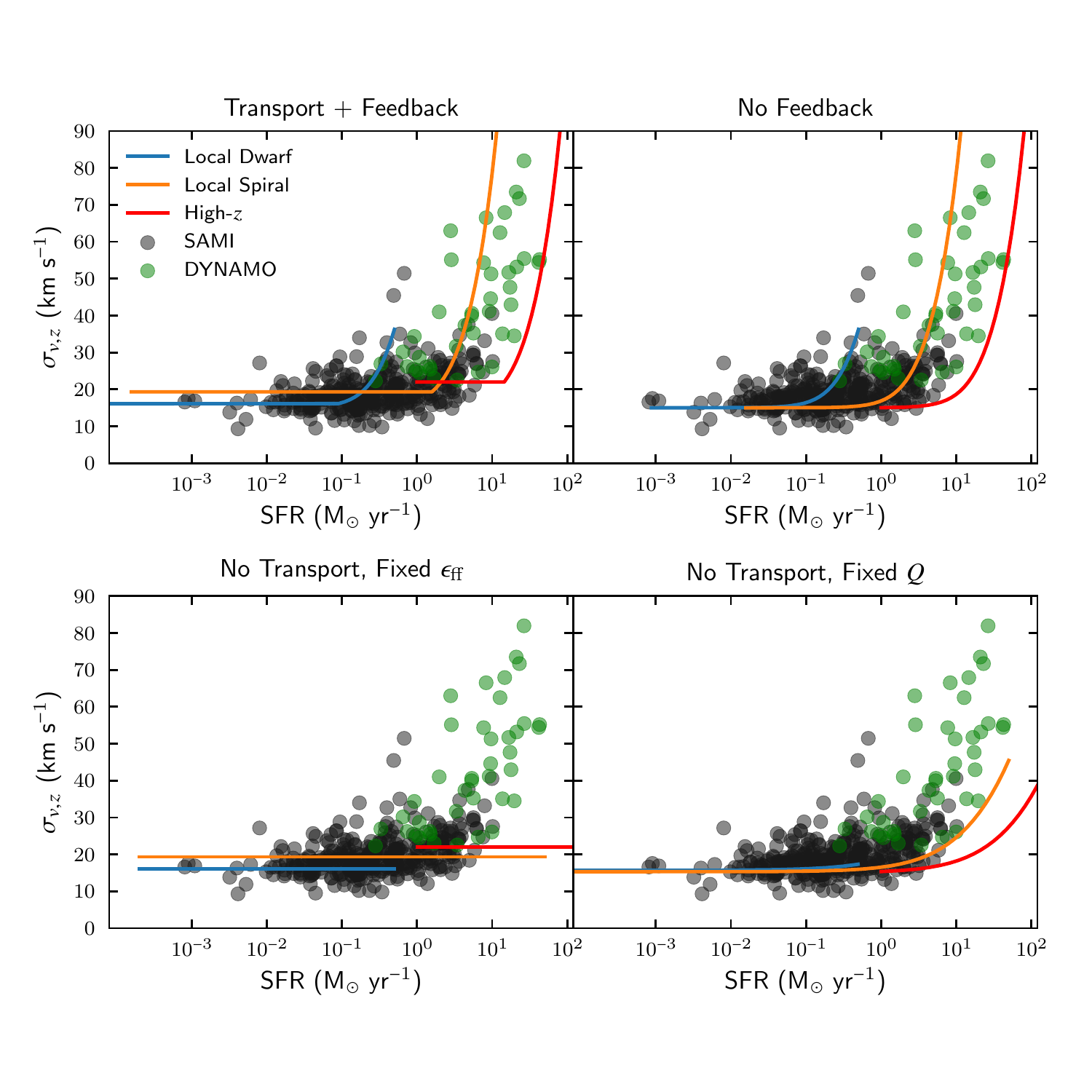} \\
    \end{center}
    \caption{Comparison of the intrinsic vertical velocity dispersion compared to the theoretical model proposed by \citet{Krumholz2018}. From left to right, we show the `Transport + Feedback', `No Feedback', `No Transport, Fixed \eff{}', and 'No Transport, Fixed $Q$' models. The individual tracks use a set of parameters (see Table \ref{tab:krumholz_params}) that represent typical galaxies for each galaxy type. We find that our observations are the most consistent with the `Transport + Feedback' model.}
    \label{fig:vdispallkrumholz2018}
\end{figure*}

We find the best agreement between our data and the `Transport + Feedback' model (Figure \ref{fig:vdispallkrumholz2018}). The lower-end of the SFR -- \sigmavz{} relation in the range SFR $\in$ [10$^{-3}$, 1] \smassyr{} is explained by the floor of the `Transport + Feedback' model tracks, which is driven by star-formation feedback processes. Importantly, the slight increase in \sigmavz{} can be explained by a change in galaxy properties across the dynamic range of SFR. The upturn in the SFR -- \sigmavz{} relation at SFR $\gtrsim$ 1 \smassyr{} is also consistent with `Transport + Feedback' model tracks. This is in contrast to the alternative models, that cannot account for the relation across the full dynamic range of SFR.

The `No Feedback' model is able to model the upturn in the SFR -- \sigmavz{} relation, but it cannot account for the lower-end of the relation. At the lower end of the relation, this model assumes \sigmavz{} approaches the thermal and expansion contributions alone. We observed that most of our galaxies lie above the assumed \mbox{\sigmavz{} > 15 \kms{}} contributions from the thermal and expansion broadening. Furthermore, there is a positive correlation of \sigmavz{} with SFR even at SFR $\lesssim$ 10 \smassyr{} that the `No Feedback' model does not appear to account for. Despite the `No Feedback' model appearing to be a better model, we note that it is difficult to distinguish between the `No Feedback model' and `Transport + Feedback' model, as the thermal and expansion broadening contribution is not well known.

The `No Transport, Fixed \eff{}' model accounts well for the lower-end SFR -- \sigmavz{} relation in our sample. However, it predicts very little evolution in \sigmavz{} across galaxy properties for low-$z$ galaxies. This is in contrast to the observations that do appear to have an upturn in \sigmavz{} for increasing SFR. This suggests that there must be an additional energetic input to the `No Transport, Fixed \eff{}' to account for increase \sigmavz{} across SFR.

The `No Transport, Fixed $Q$' model provides an alternative SFR -- \sigmavz{} relation (SFR $\propto \sigma_{v, z}^2$). The upturn in the theoretical relation qualitatively matches the observed upturn. However, the model tracks are lower than the observed \sigmavz{}. Similar to the `No Feedback' model, increasing the thermal and expansion contributions to \sigmavz{} would result in better agreement. The `No Transport, Fixed $Q$' cannot account for the increased scatter in \sigmavz{} for increasing SFR, due to estimating very little variation in \sigmavz{} across most of our dynamic range of SFR.

\begin{figure*}
    \begin{center}
        \includegraphics[
            width=\textwidth,
            keepaspectratio=true,
            trim=5mm 1mm 5mm 4mm,
            clip=true]{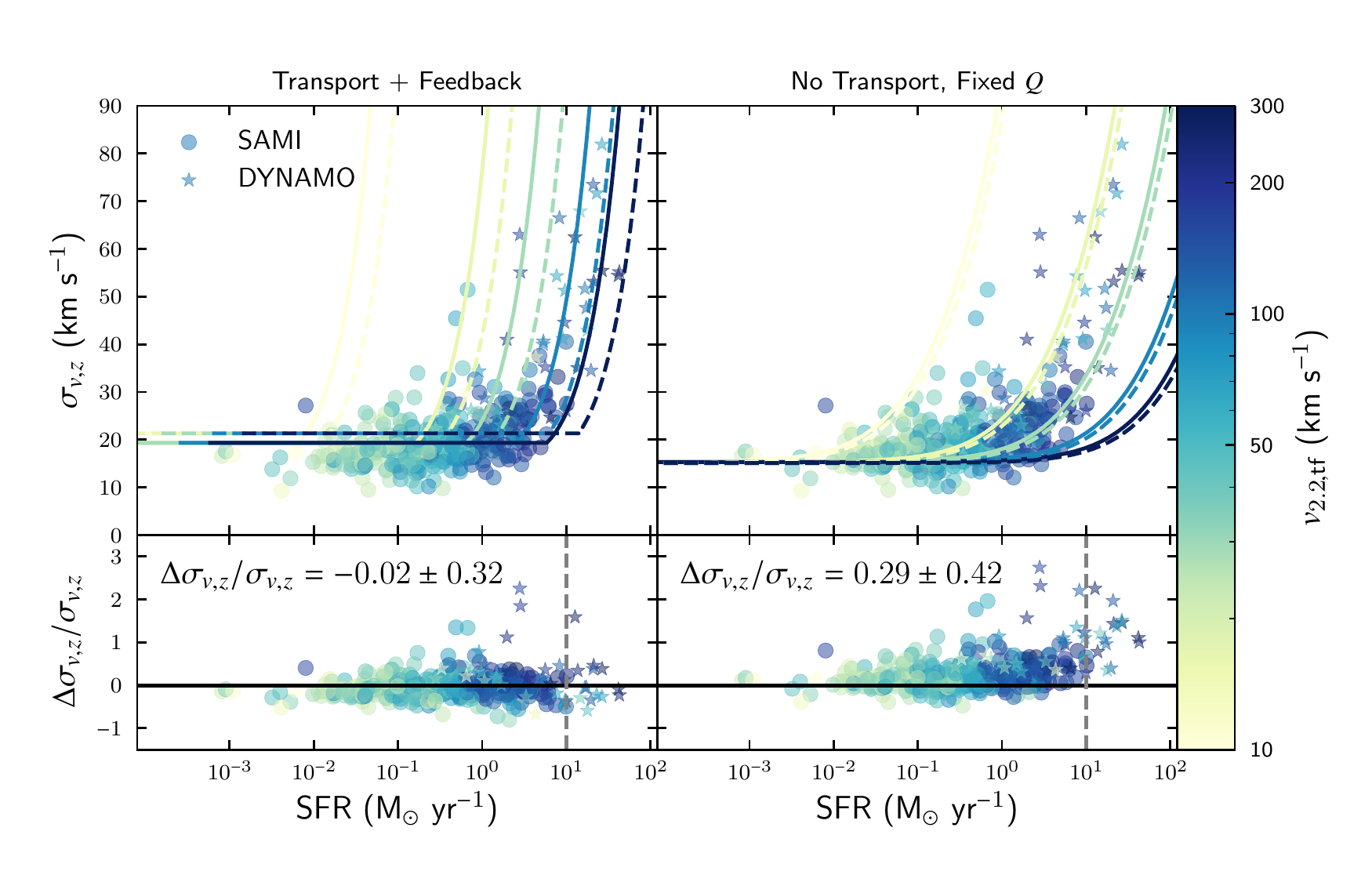} \\
    \end{center}
    \caption{Comparison of the velocity dispersions for the total galaxy sample to the `Transport + Feedback' (left) and `No Transport, Fixed $Q$' (right) models proposed by \citet{Krumholz2018}. The top two panels show the data compared to the model tracks, where the data and model tracks are colour coded by $v_{2.2, \text{tf}}$. For all other input parameters to the model tracks, the solid lines use the `local spiral'. The dashed lines use intermediate values between the `local spiral' and `high-$z$' models; $f_\text{sf} = 0.8$, $t_\text{orb} = 200$ \smassyr{}, $\beta = 0$, $f_{g, Q} = f_{g, P} = 0.6$, $\phi_a = 2$. See Table \ref{tab:krumholz_params} for the `local spiral' and `high-$z$' parameters. The bottom two panels show the relative residuals, where $\Delta \sigma_{v, z} = \sigma_{v, z} - \sigma_{v, z, \text{model}}$. We use the models represented by the solid lines for SFR < 10 \smassyr{} and the dashed lines for SFR $\geq$ 10 \smassyr{}. We also show the mean and standard deviation of the relative residuals for each model. Both theoretical models predict an increase in $\sigma_{v, z}$ as a function of SFR, however, `Transport + Feedback' provides a better fit as a function of circular velocity ($v_{2.2, \text{tf}}$).
    }
    \label{fig:vcirckrumholz2018}
\end{figure*}

To distinguish between the `Transport + Feedback' and `No Transport, Fixed $Q$' models, we also compare the theoretical model tracks while varying the circular velocity (see Figure \ref{fig:vcirckrumholz2018}). We see generally good agreement between the `Transport + Feedback' model tracks and the observed velocity dispersion. The upturn in the velocity dispersion occurs approximately at the expected circular velocity.

To quantify the differences, we calculate the relative residuals between the data and the models. To do this, we used the `local spiral' tracks for SFR < 10 \smassyr{} and a model with intermediate parameters between the `local spiral' and `high-$z$' models ($f_\text{sf} = 0.8$, $t_\text{orb} = 200$ \smassyr{}, $\beta = 0$, $f_{g, Q} = f_{g, P} = 0.6$, $\phi_a = 2$) for \mbox{SFR $\geq$ 10 \smassyr{}}. The relative residuals between the model tracks and data reveal $\Delta\sigma_{v,z}/\sigma_{v,z} = -0.02 \pm 0.32$ for the `Transport + Feedback' model compared to $\Delta\sigma_{v,z}/\sigma_{v,z} = 0.29 \pm 0.42$ for the `No Transport, Fixed $Q$' model. In particular, the relative residuals for the `No Tranport, Fixed $Q$' model increase to $\Delta\sigma_{v,z}/\sigma_{v,z} = 1.16 \pm 0.52$ for SFR > 10 \smassyr{}. Thus, suggesting that the `Transport + Feedback' model provides a better fit to the data than the `No Transport, Fixed $Q$' model.

For galaxies at SFR $\gtrsim 10$ \smassyr{}, we require a transition to values more representative of the high-$z$ galaxy model tracks, with higher $f_\text{sf}$, $f_{g, Q}$, and $f_{g, P}$ to explain the SFR -- \sigmavz{} relation. This is not surprising given that those galaxies were selected from the DYNAMO sample. Many of these galaxies exhibit similar properties to those of high-$z$ galaxies \citep{Green2014,Fisher2017} including increased molecular gas fractions \citep{Fisher2014}. 

A similar conclusion was reached by \citet{Ubler2019}, when comparing the `Transport + Feedback' model tracks as a function of circular velocity for high-$z$ galaxies. They found $\sim$60\% of their galaxies could be explained by varying the circular velocity alone.

Increasing the molecular gas fraction ($f_\text{sf}$) and the gas gravitational contribution at the mid-plane ($f_{g, Q}$, $f_{g, P}$) also shifts the base \sigmavz{} by a few \kms{}. As galaxies shift to higher $f_\text{sf}$, $f_{g, Q}$, $f_{g, P}$ as a function of SFR, this provides a mechanism to explain the increase in \sigmavz{} seen in the SAMI Galaxy Survey (see Section \ref{subsubsec:vdispcorr}).

In comparison, the `No Transport, Fixed $Q$' model predicts an increase in $\sigma_{v, z}$ as a function of SFR at a slower rate than the `Transport + Feedback' model. Comparing the model tracks when varying the circular velocity and gas properties, we find that the $\sigma_{v, z} \gtrsim 30$ \kms{} are not predicted unless assuming a much lower circular velocity ($v_{2.2, \text{tf}} \lesssim 50$ \kms{}) than expected given the stellar masses of the galaxies. Increasing the molecular gas content and gas gravitational contribution at the mid-plane as in the high-$z$ galaxies only shifts the model tracks to higher SFR.

The above analysis suggests that the `Transport + Feedback' model provides a better agreement with the data than those dominated by star-formation feedback processes. This does not completely rule out star-formation feedback processes as the primary driver, instead it may suggest that the assumed energy momentum due to star-formation feedback is too low. The assumed energy source is purely from single supernova, with momentum injection per unit of stars of \pmstar{} = 3000 \kms{}. However, \pmstar{} may be significantly higher if other sources are incorporated. For example, \citet{Gentry2017} argue that \pmstar{} could be up to an order of magnitude higher when incorporating the effects of clustered supernova. As such, further studies will be required to understand the energetic sources of star-formation feedback processes to incorporate in these models.

As a further caveat to the above analysis, we note that the theoretical models assume that we are observing the star-forming molecular gas, rather than the ionised gas. The full set of differences between the kinematics of the molecular star-forming gas compared to the ionised gas is not complete. For example, there is evidence that ionised gas may have systematically lower rotation and higher velocity dispersions compared to the molecular gas \citep{Levy2018}. However, there is limited research into these differences at this time, as such we make the assumption that these differences are minimal. Further research into the differences in molecular gas and ionised gas kinematics will be required.

\subsection{Comparing the correlation analysis to the theoretical models}

The above theoretical models (Equations \ref{eq:krumholz2018eq60} and \ref{eq:ntfq}) suggest that SFR $\propto v^2_c$, all else being set equal. Thus, we should expect a strong inverse relationship between \sigmavz{} and $v_c$. In Figure \ref{fig:globaltrends} we showed that there is a negative correlation between velocity dispersion and rotational velocity after accounting for the stellar mass contribution. We are forced to control for the stellar mass using the Tully-Fisher relation as both \sigmavz{} and $v_c$ increase for increasing stellar mass.

As such, the rotational velocity is a significant factor in prescribing the intrinsic turbulence within the galaxy. This is consistent with the theoretical models of \citet{Krumholz2018}. However, the relationship between the turbulence and rotational velocity does not distinguish between star-formation feedback or gravitational driven mechanisms of turbulence.

The proposed models also suggest a dependence of the SFR -- \sigmavz{} relation on the mid-plane gas fraction ($f_{g, P}$), the mid-plane gas contribution to the toomre-$Q$ parameter ($f_{g, Q}$), and on the molecular to neutral gas fraction ($f_\text{sf}$). \citet{Krumholz2018} also showed that galaxy turbulence driven solely by star-formation feedback has the relation SFR $\propto \sigma_v f_{g, Q}^2 / f_{g, P}$ whereas solely driven by gravitational mechanisms has SFR $\propto \sigma_v f_{g, Q}^2$.

The contribution of the gas content to the velocity dispersion is difficult to determine in our sample. We have measurements of the integrated \ion{H}{i} mass for 95 galaxies in our sample from the SAMI Galaxy Survey. We showed a slight negative but still consistent with zero correlation between the total \ion{H}{i} gas fraction ($f_{g}$) and \sigmavz{} in Section \ref{subsubsec:vdispcorr}. 

A negative correlation between integrated \ion{H}{i} mass and \sigmavz{} could be due to the expected negative correlation expected between \sigmavz{} and $f_{g, Q}$ in the `Transport + Feedback' model. However, it could also be a result of increasing  molecular gas fraction ($f_\text{sf}$) for increasing SFR and M$_*$ that are also positively correlated with \sigmavz{}. We also note that the integrated \ion{H}{i} measurements are not the ideal measurement as we cannot determine the mid-plane \ion{H}{i} gas content within each galaxy. To accurately determine the relation between \sigmavz{} and the gas content of the galaxy, we expect that resolved measurements of the \ion{H}{i} and \textsc{H$_2$}{} masses are required. In that way, we would be able to more precisely determine the mid-plane gravitational contribution of the galaxy gas content. We note that recent work by \citet[][]{Sun2020} has begun to shed light on the mid-plane gas contributions to the observed turbulence, although further studies will be required.

\section{Conclusions}
\label{sec:conclusions}

We studied the intrinsic kinematic properties of the ionised gas in 383 low-$z$ star-forming galaxies. 342 galaxies were obtained from the SAMI Galaxy Survey DR2 plus another 41 were from the DYNAMO survey. The total galaxy sample spans a wide range of galaxy properties with \mbox{SFR $\in [10^{-3}, 10^2]$ \smassyr{}}.  
The intrinsic gas kinematics were estimated using \blobby{}. \blobby{} is a flexible galaxy modelling approach that assumes that the galaxy is regularly rotating with spatially clumpy ionised gas distributions. In order to mitigate the effects of beam smearing and instrumental broadening, a convolution by the PSF and LSF on the underlying model is performed prior to calculating the likelihood function. We also performed a minor inclination correction for the sample from the SAMI Galaxy Survey to estimate the intrinsic vertical velocity dispersion (\sigmavz{}) as described in Section \ref{subsubsec:vdispcorr}. 

The sample of galaxies from the SAMI Galaxy Survey is a representation of typical galaxies at $z \lesssim 0.1$. As such, we only used that galaxy sample to determine the typical gas kinematics in galaxies at  $z \lesssim 0.1$. We find the following:
\begin{itemize}
    \item Low velocity dispersions of \mbox{\sigmavz{} $\in [14.1, 22.1]$ \kms{}} for the 68\% shortest credible interval. This is $\sim 10$ \kms{} lower than previous studies of the SAMI Galaxy Survey. The difference in results is likely driven by our beam smearing correction technique using \blobby{}, compared to the heuristic approaches applied by \citet{Zhou2017} and \citet{Johnson2018}. We also find little evidence for a significant population of galaxies with \mbox{\sigmavz{} $\gtrsim 50$ \kms{}} as found by \citet{Yu2019} in a sample of galaxies of similar galaxy properties from the MaNGA Survey. In contrast, our velocity dispersions are approximately consistent with other studies of nearby galaxies \citep{Moiseev2015,Epinat2008}.
    \item There is a significant positive correlation between \sigmavz{} and star-formation rate measures. The greatest correlation was with \sfrdensity{}. Although, the correlation is significant, the average \sigmavz{} only increased by $\sim$ 6 \kms{} for a dynamic range of \mbox{SFR $\in [10^{-3}, 10]$ \smassyr{}}. 
    \item We also find positive correlations of \sigmavz{} with integrated stellar and \ion{H}{i} gas mass as well as absolute rotational velocity.
    \item After controlling for stellar mass, there is a negative correlation between \sigmavz{} and rotational velocity. This is consistent with theoretical models proposed by \citet{Krumholz2018} for both star-formation feedback processes and gravitational driving mechanisms of turbulence.
    \item We find a weak, but still consistent with zero, negative trend between \sigmavz{} and the integrated \ion{H}{i} gas fraction. Theoretical models have suggested that there should be a relation between the gravitational contributions of the gas at the mid-plane and \sigmavz{}. We suspect that the signal between gas fraction and \sigmavz{} is lost when using the integrated \ion{H}{i} mass. Accurately determining the gravitational contributions of both \ion{H}{i} and H$_2$ at the mid-plane is likely required to observe the proposed relations.
\end{itemize}

The combined SAMI Galaxy Survey and DYNAMO data sets span a wide range of SFR, allowing for improved comparisons to the theoretical models proposed by \citet{Krumholz2018}. The SFR -- \sigmavz{} relation for our sample of galaxies is the most consistent with the `Transport + Feedback' model proposed by \citet{Krumholz2018}. We find that the SFR -- \sigmavz{} relation can be approximately explained by a transition of increasing circular velocity and molecular gas at higher SFR. 

\section*{Acknowledgements}

The SAMI Galaxy Survey is based on observations made at the Anglo-Australian Telescope. The Sydney-AAO Multi-object Integral field spectrograph (SAMI) was developed jointly by the University of Sydney and the Australian Astronomical Observatory. The SAMI input catalogue is based on data taken from the Sloan Digital Sky Survey, the GAMA Survey and the VST ATLAS Survey. The SAMI Galaxy Survey is supported by the Australian Research Council Centre of Excellence for All Sky Astrophysics in 3 Dimensions (ASTRO 3D), through project number CE170100013, the Australian Research Council Centre of Excellence for All-sky Astrophysics (CAASTRO), through project number CE110001020, and other participating institutions. The SAMI Galaxy Survey website is \url{http://sami-survey.org/}.

The authors acknowledge the University of Sydney HPC service at The University of Sydney for providing HPC and database resources that have contributed to the research results reported within this paper. 
URL: \url{http://sydney.edu.au/research_support/}

DBF and KG acknowledge support from the Australian Research Council Discovery Program grant DP160102235. DBF acknowledges support from Australian Research Council Future Fellowship FT170100376. LC is the recipient of an Australian Research Council Future Fellowship (FT180100066) funded by the Australian Government. MRK acknowledges support from Australian Research Council Future Fellowship FT180100375, and from a Humboldt Research Award from the Alexander von Humboldt Foundation. JJB acknowledges support of an Australian Research Council Future Fellowship (FT180100231). CF acknowledges funding provided by the Australian Research Council (Discovery Projects DP170100603 and Future Fellowship FT180100495), and the Australia-Germany Joint Research Cooperation Scheme (UA-DAAD). BG is the recipient of an Australian Research Council Future Fellowship (FT140101202). MSO acknowledges the funding support from the Australian Research Council through a Future Fellowship (FT140100255). JvdS is funded under JBH's ARC Laureate Fellowship (FL140100278).




\bibliographystyle{mnras}
\bibliography{main.bib} 






\onecolumn
\appendix{}
\section{Galaxy sample properties}
\label{appendix:parameters}

\begin{longtable}{lcccc ccccc cc}
    \caption{Galaxy properties for the sample from the SAMI Galaxy Survey analysed in this work. We present the ($a$) spectroscopic redshift \citep[$z_\text{spec}$,][]{Driver2011,Bryant2015}, ($b$) the stellar mass \citep[M$_*$,][]{Taylor2011,Bryant2015}, ($c$) effective radius \citep[$R_e$,][]{Taylor2011,Bryant2015}, ($d$) photometric ellipticity \citep[$e = 1 - b/a$,][]{Kelvin2012,Bryant2015}, and ($e$) SFR \citep{Gunawardhana2013,Davies2016,Driver2018} from the GAMA Survey. We also show the (f) Tully-Fisher circular velocity at $r=2.2 R_e$ calculated using the relationship proposed by \citet[$v_{2.2,\text{tf}}$,][]{Bloom2017b}. The \blobby{} inferred ($f$) circular velocity at $r=2.2 R_e$ ($v_{2.2}$) and ($g$) the LoS velocity dispersion (\sigmavlos{}). We also report the ($h$) vertical velocity dispersion (\sigmavz{}) using the inclination correction outlined in \mbox{Section \ref{subsubsec:vdispcorr}}.
    }
    \\\hline
GAMA ID & RA & Dec & $z_{\text{spec}}^a$ & log$_{10}$(M$_*$)$^b$ & $R_e^c$ &$e^d$ & log$_{10}$(SFR)$^e$ & $v_{2.2, \text{tf}}^f$ & $v_{2.2}^g$ & $\sigma_{v, \text{LoS}}^h$ & $\sigma_{v, z}^i$ \\
 & ($^\circ$) & ($^\circ$) & & (M$_\odot$) & (kpc) & & (\smassyr{}) & (\kms{}) & (\kms{}) & (\kms{}) & (\kms{}) \\
    \hline
    \endfirsthead
    \hline
GAMA ID & RA & Dec & $z_{\text{spec}}^a$ & log$_{10}$(M$_*$)$^b$ & $R_e^c$ &$e^d$ & log$_{10}$(SFR)$^e$ & $v_{2.2, \text{tf}}^f$ & $v_{2.2}^g$ & $\sigma_{v, \text{LoS}}^h$ & $\sigma_{v, z}^i$ \\
 & ($^\circ$) & ($^\circ$) & & (M$_\odot$) & (kpc) & & (\smassyr{}) & (\kms{}) & (\kms{}) & (\kms{}) & (\kms{}) \\
    \hline    
    \endhead
8353 & 182.01649 & 0.69761 & 0.020 & 9.44 & 2.43 & 0.37 & -0.35 & 99 & 112 & 22 & 19 \\
8562 & 182.79067 & 0.78576 & 0.020 & 8.42 & 2.09 & 0.28 & -1.4 & 48 & 15 & 13 & 12 \\
8570 & 182.83286 & 0.80475 & 0.021 & 9.27 & 2.27 & 0.35 & -0.84 & 88 & 60 & 18 & 16 \\
8913 & 184.22040 & 0.76587 & 0.029 & 8.79 & 1.78 & 0.45 & -4.5 & 62 & - & 29 & 25 \\
9163 & 185.14066 & 0.78806 & 0.007 & 9.22 & 2.01 & 0.45 & -1.1 & 85 & 35 & 21 & 18 \\
9352 & 185.97719 & 0.83053 & 0.024 & 8.97 & 1.12 & 0.47 & -0.83 & 71 & - & 27 & 23 \\
14555 & 212.11498 & 0.70029 & 0.026 & 8.92 & 2.54 & 0.46 & 0.18 & 68 & 72 & 21 & 17 \\
14812 & 212.93002 & 0.72011 & 0.025 & 9.99 & 2.72 & 0.24 & -0.04 & 147 & 141 & 20 & 18 \\
15218 & 214.59860 & 0.73213 & 0.026 & 9.11 & 5.22 & 0.45 & -0.68 & 78 & 115 & 22 & 19 \\
16948 & 221.10413 & 0.78286 & 0.026 & 8.89 & 2.98 & 0.17 & -0.57 & 67 & 120 & 17 & 16 \\
22932 & 179.63289 & 1.13192 & 0.039 & 9.47 & 4.06 & 0.02 & -0.28 & 101 & - & 21 & 21 \\
23337 & 181.22757 & 1.21561 & 0.021 & 9.74 & 3.02 & 0.30 & -1.0 & 123 & 98 & 16 & 14 \\
24414 & 185.53729 & 1.11275 & 0.023 & 8.35 & 2.52 & 0.30 & -1.6 & 46 & 99 & 17 & 15 \\
28654 & 211.81607 & 1.06503 & 0.035 & 9.14 & 2.38 & 0.20 & -0.56 & 80 & 102 & 21 & 19 \\
28738 & 213.15055 & 1.05790 & 0.046 & 10.05 & 2.60 & 0.42 & 0.077 & 153 & - & 27 & 23 \\
30346 & 174.63865 & -1.18449 & 0.021 & 10.45 & 5.33 & 0.32 & 0.43 & 204 & 184 & 22 & 19 \\
30377 & 174.82286 & -1.07931 & 0.027 & 8.22 & 2.29 & 0.35 & -1.4 & 42 & 96 & 20 & 17 \\
30890 & 177.25796 & -1.10260 & 0.020 & 9.79 & 3.45 & 0.43 & -0.28 & 127 & 123 & 25 & 22 \\
32249 & 183.95869 & -1.23808 & 0.021 & 8.51 & 2.72 & 0.12 & -1.2 & 51 & - & 18 & 17 \\
32274 & 184.15297 & -1.08234 & 0.021 & 8.79 & 2.18 & 0.41 & -0.89 & 62 & 74 & 19 & 16 \\
32362 & 184.53565 & -1.06411 & 0.019 & 10.41 & 6.02 & 0.44 & -0.024 & 198 & 197 & 29 & 25 \\
37050 & 215.90251 & -1.06030 & 0.031 & 9.12 & 3.75 & 0.30 & -0.7 & 79 & 100 & 19 & 17 \\
39108 & 175.13410 & -0.66962 & 0.027 & 8.35 & 1.63 & 0.17 & -0.98 & 46 & - & 25 & 23 \\
39145 & 175.43607 & -0.68800 & 0.050 & 10.20 & 2.22 & 0.24 & 0.68 & 171 & - & 42 & 38 \\
40283 & 180.46207 & -0.65541 & 0.019 & 8.90 & 3.62 & 0.23 & -1.8 & 67 & 50 & 16 & 15 \\
40420 & 181.10961 & -0.63196 & 0.020 & 9.21 & 3.62 & 0.36 & -1.6 & 84 & 108 & 25 & 22 \\
40765 & 182.89697 & -0.69958 & 0.035 & 9.04 & 0.64 & 0.41 & -0.23 & 75 & - & 41 & 35 \\
40916 & 183.54716 & -0.83157 & 0.025 & 9.82 & 6.33 & 0.45 & -0.07 & 130 & 128 & 22 & 19 \\
41173 & 184.54418 & -0.74498 & 0.021 & 8.39 & 2.25 & 0.41 & -1.4 & 47 & 23 & 18 & 15 \\
47224 & 211.86055 & -0.74540 & 0.035 & 9.16 & 1.14 & 0.40 & -0.59 & 81 & - & 19 & 17 \\
47500 & 213.25280 & -0.83100 & 0.026 & 9.49 & 1.66 & 0.46 & -0.25 & 103 & - & 26 & 22 \\
47652 & 213.60344 & -0.82934 & 0.040 & 9.43 & 2.64 & 0.14 & 0.00043 & 98 & 61 & 21 & 19 \\
49730 & 222.29648 & -0.70189 & 0.043 & 9.51 & 2.31 & 0.01 & -0.29 & 104 & - & 22 & 22 \\
49753 & 222.49249 & -0.63135 & 0.026 & 8.76 & 3.27 & 0.40 & -1.5 & 61 & 90 & 18 & 16 \\
49755 & 222.38983 & -0.78424 & 0.027 & 8.55 & 1.46 & 0.34 & -0.92 & 53 & - & 20 & 18 \\
49840 & 222.72006 & -0.67251 & 0.042 & 9.22 & 4.17 & 0.31 & -0.57 & 85 & 97 & 13 & 11 \\
53809 & 175.11901 & -0.39364 & 0.027 & 9.05 & 1.74 & 0.44 & -0.64 & 75 & - & 21 & 18 \\
53977 & 176.01840 & -0.21097 & 0.048 & 10.01 & 4.04 & 0.20 & 0.43 & 149 & 122 & 28 & 26 \\
54102 & 176.75303 & -0.29422 & 0.005 & 8.89 & 1.21 & 0.48 & -1.5 & 67 & 76 & 17 & 15 \\
54359 & 177.74299 & -0.36795 & 0.043 & 10.30 & 4.90 & 0.13 & 0.18 & 183 & - & 20 & 19 \\
54382 & 177.89815 & -0.37489 & 0.019 & 8.54 & 1.02 & 0.44 & -1.2 & 52 & - & 23 & 19 \\
54455 & 178.22625 & -0.23571 & 0.026 & 9.13 & 5.43 & 0.49 & -0.4 & 79 & 36 & 20 & 17 \\
55160 & 180.63455 & -0.38942 & 0.022 & 8.43 & 2.45 & 0.38 & -0.93 & 48 & 31 & 28 & 24 \\
55227 & 180.94630 & -0.33660 & 0.020 & 8.33 & 3.04 & 0.33 & -1.4 & 45 & 80 & 17 & 15 \\
55346 & 181.69378 & -0.27375 & 0.034 & 9.10 & 2.70 & 0.45 & -0.76 & 78 & 68 & 17 & 15 \\
55367 & 181.79334 & -0.25959 & 0.022 & 8.40 & 3.36 & 0.30 & -1.4 & 47 & 33 & 11 & 10 \\
55648 & 183.00180 & -0.37212 & 0.035 & 8.97 & 2.06 & 0.41 & -0.68 & 71 & - & 16 & 14 \\
56061 & 184.42641 & -0.22620 & 0.041 & 9.13 & 2.22 & 0.31 & -1.0 & 79 & - & 17 & 15 \\
62435 & 212.84807 & -0.30051 & 0.026 & 9.00 & 1.68 & 0.18 & -0.95 & 72 & - & 20 & 19 \\
63210 & 215.01946 & -0.31480 & 0.051 & 10.30 & 2.91 & 0.43 & -0.28 & 183 & - & 32 & 27 \\
63389 & 215.75063 & -0.25454 & 0.055 & 10.07 & 5.45 & 0.42 & 0.31 & 155 & 162 & 20 & 17 \\
63855 & 217.29079 & -0.35168 & 0.035 & 9.56 & 5.16 & 0.13 & 0.37 & 108 & - & 21 & 20 \\
64087 & 218.09196 & -0.22671 & 0.055 & 10.37 & 3.39 & 0.49 & 0.33 & 193 & - & 27 & 23 \\
65237 & 222.08657 & -0.32651 & 0.044 & 9.15 & 5.22 & 0.30 & -0.59 & 81 & 132 & 20 & 18 \\
69620 & 175.72473 & 0.16189 & 0.018 & 9.30 & 1.91 & 0.25 & -0.24 & 90 & 79 & 24 & 21 \\
69653 & 175.85485 & 0.01404 & 0.018 & 8.64 & 2.84 & 0.40 & -0.91 & 56 & 57 & 25 & 21 \\
71099 & 183.06138 & 0.07230 & 0.008 & 8.46 & 0.74 & 0.23 & -1.1 & 49 & 37 & 18 & 17 \\
71146 & 183.25125 & 0.04376 & 0.021 & 9.15 & 4.04 & 0.32 & -0.6 & 81 & 94 & 21 & 18 \\
71269 & 183.97349 & 0.08162 & 0.041 & 9.09 & 2.32 & 0.48 & 0.02 & 77 & - & 21 & 18 \\
71382 & 184.62741 & 0.01323 & 0.021 & 8.95 & 1.90 & 0.19 & -1.0 & 70 & 129 & 20 & 18 \\
77373 & 212.98003 & 0.07655 & 0.040 & 9.00 & 4.76 & 0.50 & -0.77 & 72 & 104 & 19 & 16 \\
77446 & 213.26064 & 0.14638 & 0.055 & 10.33 & 5.85 & 0.16 & 0.36 & 187 & 157 & 26 & 24 \\
77754 & 214.64775 & 0.15772 & 0.053 & 10.48 & 8.20 & 0.44 & 0.79 & 208 & 179 & 32 & 27 \\
78406 & 216.98714 & 0.02259 & 0.024 & 8.99 & 3.09 & 0.15 & -0.86 & 72 & 111 & 19 & 17 \\
78425 & 217.06865 & 0.00231 & 0.053 & 10.05 & 2.49 & 0.36 & 1.0 & 153 & - & 47 & 41 \\
78667 & 218.09082 & 0.17812 & 0.055 & 10.16 & 8.25 & 0.22 & 0.37 & 166 & 170 & 23 & 21 \\
78921 & 219.16095 & 0.11740 & 0.030 & 9.44 & 5.78 & 0.45 & -0.43 & 99 & 112 & 19 & 16 \\
79601 & 222.34769 & 0.04231 & 0.044 & 9.05 & 2.21 & 0.09 & -0.27 & 75 & - & 18 & 17 \\
79710 & 222.74198 & 0.09219 & 0.042 & 9.18 & 2.78 & 0.40 & -1.1 & 82 & 54 & 21 & 18 \\
79712 & 222.80757 & 0.02796 & 0.023 & 8.57 & 0.99 & 0.28 & -1.4 & 53 & - & 25 & 22 \\
84048 & 175.78879 & 0.55890 & 0.019 & 8.66 & 2.10 & 0.33 & -1.6 & 57 & 149 & 22 & 19 \\
84107 & 175.99843 & 0.42801 & 0.029 & 9.71 & 3.21 & 0.23 & 0.21 & 120 & 155 & 28 & 25 \\
85481 & 182.70962 & 0.59591 & 0.020 & 9.02 & 1.99 & 0.41 & -2.5 & 73 & 114 & 16 & 14 \\
86116 & 185.27934 & 0.46134 & 0.007 & 7.69 & 0.51 & 0.38 & -1.7 & 28 & - & 21 & 18 \\
91627 & 212.81851 & 0.48944 & 0.053 & 10.31 & 7.86 & 0.29 & 0.49 & 185 & 179 & 22 & 19 \\
99511 & 183.12848 & 0.89422 & 0.021 & 8.71 & 2.68 & 0.13 & -1.1 & 59 & - & 15 & 14 \\
99513 & 183.15825 & 0.89339 & 0.020 & 8.42 & 2.19 & 0.10 & -1.9 & 48 & - & 17 & 16 \\
99795 & 184.23281 & 0.91977 & 0.029 & 8.95 & 2.11 & 0.05 & -0.48 & 70 & - & 18 & 17 \\
100162 & 185.79312 & 0.93489 & 0.026 & 9.15 & 1.57 & 0.50 & -0.65 & 81 & - & 19 & 16 \\
100192 & 185.92766 & 0.96219 & 0.024 & 9.33 & 3.04 & 0.08 & -0.66 & 92 & - & 23 & 22 \\
105573 & 212.54694 & 0.86584 & 0.026 & 8.54 & 1.14 & 0.39 & -1.1 & 52 & - & 13 & 12 \\
105962 & 214.14784 & 0.88664 & 0.026 & 8.96 & 3.37 & 0.36 & -0.84 & 70 & 65 & 22 & 19 \\
106042 & 214.56214 & 0.89109 & 0.026 & 10.14 & 7.81 & 0.20 & 0.74 & 163 & 152 & 32 & 29 \\
106331 & 215.51320 & 0.86205 & 0.036 & 9.61 & 5.54 & 0.44 & -0.13 & 112 & 107 & 19 & 16 \\
106376 & 215.81121 & 0.97834 & 0.040 & 10.27 & 7.46 & 0.15 & 0.88 & 179 & 115 & 26 & 25 \\
106717 & 217.01889 & 1.00631 & 0.026 & 10.19 & 2.93 & 0.30 & 0.59 & 169 & 170 & 28 & 25 \\
107594 & 221.07590 & 0.85401 & 0.026 & 8.93 & 3.53 & 0.47 & -0.67 & 69 & 115 & 22 & 19 \\
136917 & 176.35594 & -1.73764 & 0.029 & 9.11 & 1.87 & 0.42 & -0.87 & 78 & - & 18 & 16 \\
136980 & 176.53583 & -1.82683 & 0.027 & 8.63 & 4.07 & 0.44 & -1.1 & 56 & 75 & 16 & 14 \\
137071 & 177.07578 & -1.64035 & 0.013 & 8.71 & 0.84 & 0.20 & -0.052 & 59 & - & 28 & 25 \\
137155 & 177.21879 & -1.84390 & 0.028 & 8.39 & 3.61 & 0.22 & -1.5 & 47 & 62 & 21 & 19 \\
137789 & 179.57125 & -1.72809 & 0.019 & 8.57 & 1.57 & 0.30 & -1.2 & 53 & 64 & 22 & 19 \\
137847 & 179.79836 & -1.70706 & 0.020 & 9.16 & 2.63 & 0.33 & -0.49 & 81 & 46 & 25 & 22 \\
138066 & 180.72149 & -1.77911 & 0.035 & 9.85 & 4.29 & 0.41 & -0.61 & 133 & 102 & 19 & 16 \\
138094 & 180.74242 & -1.70226 & 0.021 & 8.77 & 2.24 & 0.32 & -2.3 & 61 & 60 & 14 & 12 \\
144197 & 179.32270 & -1.37420 & 0.026 & 9.13 & 1.08 & 0.21 & -0.61 & 79 & - & 26 & 24 \\
144236 & 179.35020 & -1.31321 & 0.026 & 8.61 & 0.99 & 0.45 & -0.93 & 55 & - & 23 & 20 \\
144320 & 179.73348 & -1.43043 & 0.052 & 10.27 & 1.96 & 0.30 & -0.03 & 179 & - & 34 & 30 \\
144402 & 179.96120 & -1.38195 & 0.036 & 10.25 & 3.25 & 0.35 & 0.55 & 177 & 172 & 35 & 31 \\
144497 & 180.37719 & -1.43612 & 0.035 & 9.28 & 1.09 & 0.12 & -0.17 & 88 & - & 55 & 51 \\
144682 & 181.03465 & -1.41719 & 0.035 & 9.02 & 1.03 & 0.41 & -0.77 & 73 & - & 40 & 34 \\
145267 & 183.70061 & -1.34594 & 0.032 & 9.12 & 1.37 & 0.47 & -1.1 & 79 & - & 31 & 26 \\
145583 & 185.32451 & -1.25413 & 0.022 & 9.39 & 3.61 & 0.41 & -0.8 & 96 & 85 & 17 & 14 \\
176955 & 174.94289 & -1.87526 & 0.058 & 10.62 & 9.09 & 0.34 & 0.7 & 230 & 204 & 21 & 18 \\
177081 & 175.53937 & -1.90905 & 0.020 & 8.92 & 1.47 & 0.33 & -0.34 & 68 & 81 & 30 & 26 \\
177481 & 176.91006 & -1.92285 & 0.027 & 8.84 & 1.61 & 0.30 & -1.4 & 65 & - & 22 & 19 \\
178481 & 180.44250 & -1.93475 & 0.025 & 9.00 & 4.51 & 0.29 & -0.72 & 72 & 109 & 20 & 18 \\
178580 & 180.81309 & -1.95678 & 0.021 & 8.43 & 1.75 & 0.00 & -1.4 & 48 & - & 20 & 20 \\
183932 & 174.27021 & -1.60977 & 0.022 & 8.27 & 1.86 & 0.16 & -1.2 & 43 & 26 & 21 & 20 \\
184234 & 175.68429 & -1.48754 & 0.029 & 9.01 & 4.37 & 0.05 & -0.62 & 73 & - & 21 & 21 \\
184370 & 176.21728 & -1.53212 & 0.026 & 9.65 & 2.55 & 0.14 & -0.56 & 115 & 36 & 20 & 19 \\
184415 & 176.34198 & -1.56521 & 0.028 & 9.56 & 2.28 & 0.24 & -0.26 & 108 & 111 & 20 & 18 \\
185190 & 179.49465 & -1.55768 & 0.020 & 9.01 & 2.45 & 0.32 & -0.86 & 73 & 83 & 16 & 14 \\
185252 & 179.54589 & -1.64745 & 0.022 & 8.46 & 3.57 & 0.39 & -1.7 & 49 & 48 & 21 & 18 \\
185291 & 179.80472 & -1.60447 & 0.022 & 8.83 & 2.38 & 0.41 & -1.1 & 64 & 53 & 24 & 21 \\
185532 & 180.69427 & -1.59343 & 0.020 & 9.28 & 3.45 & 0.12 & -0.91 & 88 & - & 17 & 16 \\
185557 & 180.75343 & -1.63802 & 0.019 & 9.62 & 1.13 & 0.24 & - & 113 & - & 23 & 21 \\
185622 & 181.08444 & -1.53028 & 0.005 & 7.87 & 6.13 & 0.29 & -2.4 & 32 & 43 & 11 & 9 \\
197419 & 135.20729 & -0.71429 & 0.041 & 9.30 & 3.62 & 0.40 & -0.58 & 90 & 103 & 17 & 14 \\
198503 & 139.76575 & -0.81766 & 0.017 & 8.58 & 0.93 & 0.46 & -1.3 & 54 & - & 29 & 25 \\
198817 & 140.97499 & -0.68263 & 0.055 & 10.09 & 4.75 & 0.20 & 0.16 & 158 & 184 & 24 & 22 \\
203148 & 132.84017 & -0.39516 & 0.043 & 9.27 & 1.77 & 0.12 & -0.26 & 88 & - & 26 & 25 \\
203684 & 134.79005 & -0.27214 & 0.042 & 9.19 & 3.54 & 0.46 & -0.48 & 83 & 117 & 18 & 15 \\
203729 & 135.04616 & -0.30183 & 0.042 & 9.44 & 2.17 & 0.44 & -0.31 & 99 & - & 53 & 45 \\
203998 & 136.14023 & -0.31481 & 0.028 & 8.93 & 1.60 & 0.09 & -0.78 & 69 & - & 16 & 15 \\
204096 & 136.52107 & -0.26037 & 0.040 & 9.98 & 3.82 & 0.17 & 0.0099 & 146 & 158 & 16 & 15 \\
204868 & 139.84670 & -0.21330 & 0.039 & 9.49 & 1.10 & 0.19 & -0.27 & 103 & - & 21 & 20 \\
208520 & 129.40912 & 0.05067 & 0.035 & 9.65 & 4.74 & 0.16 & -0.45 & 115 & 121 & 17 & 15 \\
208892 & 130.75455 & 0.16933 & 0.029 & 9.39 & 7.04 & 0.48 & -0.81 & 96 & 100 & 17 & 15 \\
209181 & 132.12520 & 0.17087 & 0.058 & 10.30 & 5.53 & 0.23 & 0.79 & 183 & 173 & 30 & 27 \\
209414 & 133.20974 & 0.15797 & 0.026 & 9.04 & 3.84 & 0.45 & -1.0 & 75 & 90 & 24 & 20 \\
209743 & 134.67676 & 0.19143 & 0.041 & 10.16 & 6.20 & 0.48 & 0.018 & 166 & 180 & 18 & 16 \\
210060 & 136.40777 & 0.00327 & 0.019 & 8.98 & 4.92 & 0.21 & -1.0 & 71 & 99 & 17 & 15 \\
210567 & 138.74414 & 0.20803 & 0.057 & 9.48 & 5.79 & 0.13 & -0.37 & 102 & - & 14 & 14 \\
210781 & 139.64824 & 0.05988 & 0.055 & 10.22 & 5.31 & 0.24 & -0.089 & 173 & 168 & 18 & 16 \\
210808 & 139.75689 & 0.17252 & 0.017 & 8.41 & 1.47 & 0.13 & -1.9 & 48 & 95 & 15 & 14 \\
210909 & 140.28626 & 0.08058 & 0.024 & 8.44 & 1.73 & 0.49 & -1.9 & 49 & 55 & 23 & 20 \\
214245 & 129.52446 & 0.60896 & 0.014 & 9.40 & 1.35 & 0.32 & -1.6 & 96 & 75 & 20 & 17 \\
214860 & 131.89667 & 0.56184 & 0.058 & 9.75 & 7.03 & 0.49 & 0.003 & 124 & 104 & 25 & 21 \\
216843 & 140.19242 & 0.60472 & 0.024 & 9.26 & 4.06 & 0.29 & -0.68 & 87 & 93 & 21 & 18 \\
220275 & 180.92608 & 1.45729 & 0.021 & 9.14 & 2.69 & 0.02 & -0.89 & 80 & - & 17 & 17 \\
220319 & 180.99245 & 1.48278 & 0.021 & 8.57 & 2.33 & 0.21 & -1.8 & 53 & 24 & 16 & 14 \\
220371 & 181.23715 & 1.50824 & 0.020 & 9.53 & 3.23 & 0.35 & -0.93 & 106 & 134 & 21 & 18 \\
220372 & 181.28939 & 1.55929 & 0.021 & 9.06 & 1.86 & 0.12 & -1.3 & 76 & - & 18 & 17 \\
220439 & 181.63159 & 1.61663 & 0.019 & 9.54 & 2.54 & 0.18 & -0.26 & 107 & 134 & 14 & 13 \\
220578 & 182.17817 & 1.45636 & 0.019 & 8.98 & 1.28 & 0.41 & -0.78 & 71 & - & 17 & 14 \\
220687 & 182.83299 & 1.49227 & 0.007 & 9.27 & 3.36 & 0.43 & -0.75 & 88 & 74 & 17 & 15 \\
220750 & 182.98977 & 1.48925 & 0.021 & 8.62 & 2.42 & 0.30 & -0.87 & 55 & 70 & 16 & 14 \\
221369 & 185.83472 & 1.61648 & 0.027 & 8.64 & 1.26 & 0.34 & -0.56 & 56 & - & 23 & 20 \\
227036 & 211.82817 & 1.28196 & 0.035 & 9.56 & 4.02 & 0.39 & 0.19 & 108 & 136 & 25 & 22 \\
227223 & 212.67106 & 1.33941 & 0.055 & 10.31 & 4.44 & 0.11 & 0.75 & 185 & - & 31 & 29 \\
227289 & 212.82231 & 1.35262 & 0.026 & 9.17 & 4.75 & 0.08 & -0.68 & 82 & - & 21 & 20 \\
227673 & 214.53595 & 1.22412 & 0.026 & 9.35 & 3.15 & 0.13 & -0.28 & 93 & - & 24 & 23 \\
227970 & 215.60459 & 1.19760 & 0.054 & 10.16 & 5.19 & 0.24 & 0.47 & 166 & 174 & 16 & 15 \\
228086 & 216.08084 & 1.12442 & 0.039 & 9.18 & 4.21 & 0.20 & -0.38 & 82 & 60 & 19 & 17 \\
230174 & 178.74753 & 1.85812 & 0.021 & 8.48 & 1.85 & 0.19 & -1.9 & 50 & 51 & 18 & 17 \\
238328 & 213.96582 & 1.58638 & 0.025 & 8.82 & 1.36 & 0.36 & -1.3 & 64 & - & 20 & 17 \\
238395 & 214.24319 & 1.64043 & 0.025 & 9.87 & 2.26 & 0.18 & 0.28 & 135 & 110 & 31 & 29 \\
238406 & 214.20244 & 1.75963 & 0.056 & 10.45 & 8.32 & 0.37 & 0.25 & 204 & 194 & 27 & 24 \\
239490 & 217.99757 & 1.58140 & 0.030 & 9.21 & 3.68 & 0.19 & -0.58 & 84 & 68 & 21 & 19 \\
240108 & 220.62338 & 1.50040 & 0.007 & 9.02 & 1.20 & 0.42 & -1.3 & 73 & 75 & 20 & 17 \\
240202 & 221.12828 & 1.52201 & 0.005 & 8.66 & 1.29 & 0.21 & -2.0 & 57 & 41 & 17 & 15 \\
250277 & 214.43384 & 1.98131 & 0.058 & 10.01 & 5.72 & 0.29 & 0.23 & 149 & 35 & 31 & 28 \\
251297 & 218.11956 & 1.91052 & 0.030 & 9.52 & 4.10 & 0.30 & -0.31 & 105 & 116 & 18 & 16 \\
251367 & 218.23409 & 1.89580 & 0.030 & 9.04 & 2.25 & 0.30 & -0.85 & 75 & 87 & 21 & 19 \\
252074 & 221.96823 & 1.80223 & 0.028 & 8.58 & 3.14 & 0.42 & -1.3 & 54 & 30 & 18 & 15 \\
271562 & 174.75468 & 1.33657 & 0.005 & 7.82 & 0.70 & 0.41 & -0.81 & 31 & 31 & 24 & 20 \\
272996 & 181.66757 & 1.33397 & 0.022 & 8.76 & 1.88 & 0.49 & -1.3 & 61 & 58 & 21 & 17 \\
273092 & 181.99998 & 1.39593 & 0.037 & 10.07 & 6.27 & 0.25 & 0.19 & 155 & 37 & 20 & 18 \\
273242 & 182.79525 & 1.44168 & 0.019 & 8.68 & 2.78 & 0.15 & -1.3 & 58 & 83 & 19 & 18 \\
273296 & 182.99771 & 1.35004 & 0.021 & 9.56 & 5.31 & 0.46 & -0.28 & 108 & 68 & 19 & 16 \\
273309 & 183.03839 & 1.31149 & 0.020 & 9.24 & 3.60 & 0.11 & -1.1 & 86 & - & 19 & 18 \\
273951 & 185.93037 & 1.31109 & 0.026 & 8.72 & 2.53 & 0.45 & -0.38 & 59 & 7 & 28 & 24 \\
278074 & 211.96000 & 1.13692 & 0.025 & 9.78 & 4.56 & 0.17 & -1.4 & 126 & 60 & 15 & 14 \\
278554 & 132.30501 & 0.78322 & 0.043 & 9.00 & 4.05 & 0.09 & -0.68 & 72 & - & 19 & 18 \\
278684 & 133.13103 & 0.85357 & 0.011 & 8.09 & 0.48 & 0.20 & -1.8 & 38 & - & 24 & 22 \\
278804 & 133.85939 & 0.85818 & 0.042 & 9.82 & 2.45 & 0.38 & -0.97 & 130 & - & 15 & 13 \\
278909 & 134.42490 & 0.81731 & 0.041 & 9.33 & 2.37 & 0.48 & -0.78 & 92 & - & 15 & 13 \\
279066 & 135.13286 & 0.97642 & 0.018 & 8.25 & 4.71 & 0.28 & - & 42 & 10 & 16 & 15 \\
279818 & 139.43876 & 1.05542 & 0.027 & 9.54 & 4.35 & 0.21 & -0.26 & 107 & 42 & 22 & 20 \\
279917 & 139.99533 & 0.96084 & 0.018 & 9.32 & 3.88 & 0.44 & -0.34 & 91 & 76 & 24 & 20 \\
289107 & 181.04059 & 1.82596 & 0.017 & 9.68 & 3.73 & 0.36 & -0.75 & 118 & 168 & 15 & 13 \\
296639 & 212.67738 & 1.40807 & 0.046 & 10.22 & 3.13 & 0.18 & 0.17 & 173 & 132 & 22 & 20 \\
296742 & 213.20535 & 1.48923 & 0.018 & 9.15 & 1.38 & 0.46 & -1.1 & 81 & 29 & 31 & 26 \\
296934 & 214.04425 & 1.54141 & 0.053 & 10.21 & 3.82 & 0.20 & 0.28 & 172 & 174 & 25 & 23 \\
297633 & 216.56453 & 1.49149 & 0.055 & 10.43 & 6.88 & 0.25 & 0.38 & 201 & 175 & 19 & 17 \\
297694 & 216.86676 & 1.33773 & 0.025 & 9.11 & 12.31 & 0.14 & -1.6 & 78 & 195 & 15 & 14 \\
298114 & 218.40091 & 1.30590 & 0.056 & 10.25 & 5.93 & 0.41 & 0.49 & 177 & 175 & 24 & 20 \\
298738 & 221.59337 & 1.22840 & 0.050 & 10.06 & 5.64 & 0.43 & 0.15 & 154 & 154 & 22 & 19 \\
300350 & 129.16480 & 1.13610 & 0.014 & 8.32 & 1.49 & 0.22 & -3.1 & 45 & 41 & 18 & 17 \\
300372 & 129.29410 & 1.00136 & 0.039 & 9.16 & 1.23 & 0.18 & -0.86 & 81 & - & 26 & 24 \\
300477 & 129.70677 & 1.12101 & 0.029 & 9.25 & 3.53 & 0.26 & -0.64 & 87 & 130 & 22 & 20 \\
300787 & 130.93495 & 1.07919 & 0.044 & 10.32 & 2.97 & 0.32 & 0.045 & 186 & 199 & 15 & 13 \\
300821 & 131.03734 & 1.21435 & 0.013 & 8.82 & 0.98 & 0.29 & -0.64 & 64 & 88 & 23 & 20 \\
301346 & 133.52459 & 1.19186 & 0.044 & 10.16 & 3.33 & 0.46 & 0.42 & 166 & 170 & 28 & 24 \\
301885 & 135.53948 & 1.22605 & 0.057 & 10.60 & 11.86 & 0.40 & 0.56 & 227 & 263 & 21 & 18 \\
318936 & 212.94107 & 1.94731 & 0.018 & 8.90 & 2.77 & 0.36 & -0.61 & 67 & 100 & 22 & 19 \\
319150 & 213.62262 & 1.81263 & 0.025 & 8.56 & 1.06 & 0.38 & -1.0 & 53 & - & 23 & 20 \\
320068 & 216.87191 & 1.85175 & 0.029 & 9.17 & 2.38 & 0.28 & -0.59 & 82 & 102 & 22 & 20 \\
320281 & 217.63635 & 1.85328 & 0.034 & 9.84 & 3.02 & 0.39 & -0.11 & 132 & 185 & 26 & 23 \\
322910 & 129.39531 & 1.57389 & 0.031 & 9.74 & 3.61 & 0.19 & -0.48 & 123 & 30 & 25 & 23 \\
323194 & 130.81630 & 1.48410 & 0.013 & 8.61 & 0.81 & 0.37 & -1.4 & 55 & - & 17 & 15 \\
323224 & 130.98705 & 1.58429 & 0.013 & 8.61 & 1.02 & 0.11 & -0.78 & 55 & - & 18 & 17 \\
323242 & 131.00309 & 1.67133 & 0.028 & 9.50 & 1.38 & 0.23 & -0.4 & 104 & - & 36 & 33 \\
323504 & 131.95082 & 1.53447 & 0.063 & 10.94 & 11.71 & 0.09 & 0.41 & 289 & - & 27 & 26 \\
323507 & 132.03504 & 1.56604 & 0.040 & 9.44 & 2.92 & 0.14 & -0.3 & 99 & 144 & 23 & 21 \\
323874 & 133.49341 & 1.66407 & 0.058 & 10.56 & 4.42 & 0.08 & -0.61 & 221 & - & 17 & 17 \\
324323 & 135.50044 & 1.78604 & 0.053 & 9.74 & 2.64 & 0.48 & -0.63 & 123 & - & 12 & 10 \\
325533 & 140.92832 & 2.00336 & 0.053 & 10.10 & 6.79 & 0.48 & -0.12 & 159 & 172 & 17 & 15 \\
345646 & 130.40960 & 1.96809 & 0.014 & 8.44 & 3.75 & 0.08 & -1.2 & 49 & - & 25 & 24 \\
346257 & 133.04215 & 1.98304 & 0.029 & 8.63 & 1.21 & 0.50 & -1.1 & 56 & - & 22 & 19 \\
346440 & 133.74686 & 2.13436 & 0.020 & 8.37 & 0.50 & 0.11 & -1.2 & 46 & - & 19 & 18 \\
346718 & 134.86958 & 2.06157 & 0.057 & 9.46 & 1.77 & 0.19 & 0.33 & 101 & - & 27 & 25 \\
346861 & 135.29644 & 2.07820 & 0.055 & 9.82 & 5.15 & 0.33 & -0.26 & 130 & 132 & 16 & 14 \\
347263 & 136.99176 & 2.27055 & 0.026 & 9.48 & 2.92 & 0.46 & -0.55 & 102 & 184 & 22 & 19 \\
375904 & 131.27015 & 1.40141 & 0.014 & 8.07 & 1.06 & 0.35 & -1.5 & 37 & 51 & 21 & 18 \\
376165 & 132.17024 & 1.49956 & 0.029 & 8.70 & 2.94 & 0.12 & -1.4 & 58 & - & 19 & 18 \\
376185 & 132.35194 & 1.38501 & 0.034 & 9.07 & 2.16 & 0.30 & -0.75 & 76 & - & 21 & 19 \\
377348 & 137.33399 & 1.61430 & 0.004 & 7.59 & 0.44 & 0.17 & -2.2 & 26 & 53 & 19 & 17 \\
378060 & 140.38950 & 1.58462 & 0.017 & 8.70 & 3.10 & 0.34 & -1.6 & 58 & 56 & 19 & 17 \\
382152 & 135.42424 & 1.85215 & 0.057 & 10.12 & 6.77 & 0.28 & -0.22 & 161 & 120 & 24 & 21 \\
382631 & 137.71356 & 2.02189 & 0.055 & 10.09 & 2.92 & 0.16 & -0.015 & 158 & - & 19 & 17 \\
382764 & 138.26745 & 2.03871 & 0.013 & 9.05 & 1.18 & 0.34 & -0.43 & 75 & 138 & 30 & 26 \\
383033 & 139.59121 & 2.17020 & 0.027 & 8.47 & 2.22 & 0.31 & -1.4 & 50 & 49 & 20 & 18 \\
383259 & 140.67041 & 2.11154 & 0.057 & 10.73 & 6.74 & 0.42 & 0.9 & 249 & 143 & 39 & 33 \\
383318 & 140.95009 & 2.11275 & 0.024 & 9.92 & 5.05 & 0.48 & -0.085 & 140 & 66 & 39 & 33 \\
386286 & 131.34372 & 2.19006 & 0.006 & 8.22 & 1.11 & 0.24 & -1.6 & 42 & 52 & 20 & 18 \\
386898 & 134.40439 & 2.23945 & 0.054 & 10.44 & 8.66 & 0.21 & 0.31 & 202 & 225 & 17 & 16 \\
388603 & 140.78384 & 2.48607 & 0.017 & 9.80 & 5.50 & 0.12 & -0.4 & 128 & - & 22 & 20 \\
418624 & 137.09716 & 2.54414 & 0.055 & 10.01 & 4.72 & 0.30 & -0.15 & 149 & 176 & 20 & 18 \\
418795 & 137.76469 & 2.57229 & 0.039 & 9.13 & 1.62 & 0.24 & -0.77 & 79 & - & 11 & 10 \\
419632 & 140.75064 & 2.86863 & 0.025 & 8.85 & 1.80 & 0.13 & -3.0 & 65 & 17 & 19 & 18 \\
422355 & 130.50504 & 2.52837 & 0.028 & 9.26 & 3.18 & 0.04 & -0.63 & 87 & - & 19 & 19 \\
422359 & 130.55488 & 2.62461 & 0.050 & 10.07 & 2.72 & 0.45 & 0.14 & 155 & - & 14 & 12 \\
422366 & 130.59560 & 2.49733 & 0.029 & 9.62 & 5.58 & 0.49 & -0.35 & 113 & 90 & 22 & 18 \\
422486 & 131.18034 & 2.57274 & 0.026 & 8.78 & 2.15 & 0.42 & -0.79 & 62 & 135 & 21 & 18 \\
422619 & 131.78175 & 2.62180 & 0.029 & 9.63 & 4.52 & 0.14 & -1.8 & 114 & 97 & 17 & 16 \\
463660 & 213.92314 & -1.15695 & 0.038 & 9.02 & 3.18 & 0.08 & -1.4 & 73 & - & 15 & 14 \\
485504 & 216.10103 & -1.76490 & 0.056 & 10.20 & 6.57 & 0.23 & 0.21 & 171 & 184 & 18 & 16 \\
485529 & 216.24765 & -1.86856 & 0.030 & 9.07 & 1.83 & 0.46 & -0.36 & 76 & - & 30 & 25 \\
485834 & 217.57879 & -1.78770 & 0.056 & 10.69 & 6.41 & 0.43 & 0.54 & 242 & 250 & 29 & 24 \\
485885 & 217.75790 & -1.71721 & 0.055 & 10.25 & 6.09 & 0.16 & 0.76 & 177 & 167 & 23 & 21 \\
487010 & 222.52592 & -1.61157 & 0.043 & 9.01 & 2.93 & 0.19 & -0.9 & 73 & 67 & 17 & 16 \\
487027 & 222.67911 & -1.71488 & 0.026 & 10.09 & 3.58 & 0.35 & 0.57 & 158 & 149 & 33 & 28 \\
487175 & 223.33977 & -1.59495 & 0.042 & 9.73 & 3.48 & 0.28 & 0.32 & 122 & 127 & 24 & 22 \\
492384 & 216.39461 & -1.37612 & 0.055 & 10.46 & 4.59 & 0.45 & 0.12 & 205 & 172 & 31 & 26 \\
492414 & 216.50320 & -1.41180 & 0.055 & 10.10 & 5.31 & 0.02 & 0.33 & 159 & - & 23 & 23 \\
493621 & 221.83561 & -1.30299 & 0.029 & 9.03 & 3.37 & 0.23 & -1.3 & 74 & 109 & 20 & 18 \\
493812 & 222.52657 & -1.16131 & 0.043 & 9.54 & 4.59 & 0.45 & -0.97 & 107 & 136 & 26 & 22 \\
493825 & 222.43912 & -1.17427 & 0.027 & 8.23 & 1.91 & 0.39 & -1.3 & 42 & 51 & 21 & 18 \\
508421 & 216.98916 & -1.63118 & 0.055 & 10.39 & 4.55 & 0.26 & -0.2 & 195 & 192 & 25 & 22 \\
508680 & 217.90221 & -1.59247 & 0.030 & 9.25 & 3.54 & 0.25 & -0.51 & 87 & 93 & 16 & 14 \\
509397 & 221.19366 & -1.51910 & 0.056 & 10.24 & 6.82 & 0.19 & -0.053 & 176 & 109 & 20 & 19 \\
509444 & 221.32078 & -1.56930 & 0.034 & 9.05 & 3.65 & 0.35 & -1.2 & 75 & 88 & 21 & 18 \\
509557 & 221.96775 & -1.57005 & 0.027 & 8.87 & 0.56 & 0.38 & -0.81 & 66 & - & 33 & 29 \\
509576 & 221.97272 & -1.37673 & 0.027 & 8.26 & 2.72 & 0.32 & -0.65 & 43 & 66 & 20 & 17 \\
509670 & 222.34731 & -1.55925 & 0.027 & 8.95 & 4.31 & 0.43 & -0.71 & 70 & 118 & 22 & 18 \\
509852 & 223.13292 & -1.34509 & 0.043 & 10.07 & 7.69 & 0.34 & 0.29 & 155 & 105 & 19 & 17 \\
511867 & 216.38846 & -1.11394 & 0.055 & 10.68 & 7.47 & 0.40 & 1.0 & 240 & 218 & 32 & 28 \\
511921 & 216.67460 & -1.14927 & 0.031 & 9.16 & 1.19 & 0.38 & -0.46 & 81 & - & 24 & 21 \\
512524 & 219.06927 & -1.13120 & 0.040 & 9.27 & 5.47 & 0.08 & -0.59 & 88 & - & 16 & 16 \\
513108 & 221.71563 & -1.14686 & 0.042 & 9.64 & 7.81 & 0.14 & -0.5 & 114 & 135 & 25 & 24 \\
514029 & 214.13351 & -1.18215 & 0.050 & 10.49 & 6.38 & 0.30 & 0.51 & 210 & 187 & 19 & 17 \\
517167 & 131.16137 & 2.41098 & 0.030 & 9.24 & 2.41 & 0.31 & -0.12 & 86 & 114 & 19 & 17 \\
517249 & 131.55101 & 2.41047 & 0.028 & 9.40 & 3.38 & 0.38 & -0.29 & 96 & 83 & 24 & 21 \\
517306 & 131.71344 & 2.56971 & 0.030 & 9.38 & 3.07 & 0.37 & -0.34 & 95 & 90 & 22 & 19 \\
517960 & 134.27689 & 2.66458 & 0.013 & 8.30 & 2.59 & 0.39 & -1.4 & 44 & 28 & 17 & 15 \\
521736 & 130.67894 & 2.87319 & 0.050 & 9.87 & 1.71 & 0.33 & 0.57 & 135 & - & 40 & 35 \\
521768 & 131.07263 & 2.88117 & 0.050 & 10.19 & 4.78 & 0.22 & 0.036 & 169 & 160 & 22 & 20 \\
521894 & 131.65458 & 2.82703 & 0.013 & 8.77 & 1.58 & 0.26 & -1.5 & 61 & 106 & 19 & 17 \\
521898 & 131.68612 & 2.79428 & 0.028 & 8.46 & 1.19 & 0.09 & -1.1 & 49 & - & 22 & 21 \\
534654 & 174.35287 & -0.96382 & 0.050 & 10.31 & 3.84 & 0.03 & -0.0073 & 185 & - & 22 & 22 \\
534753 & 175.02585 & -0.90142 & 0.029 & 10.35 & 1.14 & 0.33 & 0.22 & 190 & - & 27 & 24 \\
535283 & 177.25575 & -0.88835 & 0.020 & 8.50 & 0.72 & 0.50 & -1.3 & 51 & - & 19 & 16 \\
535974 & 179.96350 & -0.85869 & 0.036 & 9.27 & 1.96 & 0.40 & -0.28 & 88 & - & 21 & 18 \\
537399 & 185.08379 & -0.88202 & 0.040 & 9.63 & 4.51 & 0.32 & -0.33 & 114 & 116 & 18 & 16 \\
537476 & 185.39249 & -1.00951 & 0.021 & 8.21 & 1.49 & 0.42 & -0.56 & 41 & 23 & 25 & 22 \\
543752 & 212.63639 & -0.84186 & 0.025 & 8.92 & 5.06 & 0.47 & -1.3 & 68 & 53 & 23 & 19 \\
543763 & 212.75337 & -0.90393 & 0.026 & 8.49 & 2.01 & 0.49 & -1.8 & 50 & 76 & 20 & 17 \\
543860 & 213.15467 & -1.01222 & 0.054 & 10.01 & 4.94 & 0.20 & 0.31 & 149 & 44 & 31 & 28 \\
544084 & 213.89591 & -1.03869 & 0.038 & 9.04 & 5.12 & 0.42 & -0.81 & 75 & 169 & 17 & 15 \\
544812 & 216.98074 & -1.00818 & 0.029 & 9.32 & 3.42 & 0.41 & -1.1 & 91 & 116 & 17 & 15 \\
544853 & 217.37900 & -0.88385 & 0.035 & 9.54 & 7.30 & 0.18 & -1.4 & 107 & 96 & 16 & 15 \\
546043 & 222.74183 & -0.88154 & 0.027 & 9.43 & 3.01 & 0.25 & -0.36 & 98 & 104 & 21 & 19 \\
551192 & 139.33882 & -0.45421 & 0.017 & 8.75 & 0.61 & 0.32 & -1.0 & 61 & - & 33 & 29 \\
551368 & 140.01779 & -0.50248 & 0.026 & 8.88 & 1.09 & 0.04 & -1.2 & 66 & - & 24 & 23 \\
558887 & 174.37404 & -0.47340 & 0.029 & 8.83 & 2.28 & 0.40 & -0.97 & 64 & 77 & 21 & 18 \\
559292 & 176.41768 & -0.57082 & 0.028 & 8.71 & 2.52 & 0.47 & -0.58 & 59 & 41 & 26 & 22 \\
559300 & 176.53218 & -0.45799 & 0.013 & 8.64 & 1.23 & 0.23 & -1.7 & 56 & 46 & 18 & 16 \\
559495 & 177.34230 & -0.62371 & 0.040 & 9.05 & 3.20 & 0.43 & -0.18 & 75 & 138 & 21 & 18 \\
560333 & 179.98443 & -0.54822 & 0.022 & 9.90 & 5.52 & 0.18 & -0.18 & 138 & 154 & 17 & 16 \\
560718 & 181.33290 & -0.48020 & 0.005 & 7.76 & 0.50 & 0.14 & -3.0 & 30 & 63 & 18 & 17 \\
560946 & 182.33179 & -0.52747 & 0.035 & 9.19 & 3.44 & 0.35 & -0.97 & 83 & 88 & 13 & 12 \\
561143 & 183.01351 & -0.60685 & 0.035 & 9.54 & 4.33 & 0.47 & 0.15 & 107 & 64 & 37 & 31 \\
567676 & 212.76660 & -0.54511 & 0.026 & 8.57 & 2.42 & 0.17 & -1.8 & 53 & 21 & 23 & 21 \\
567736 & 213.05273 & -0.61270 & 0.025 & 8.69 & 1.81 & 0.47 & 0.32 & 58 & 15 & 28 & 23 \\
567760 & 213.07541 & -0.56930 & 0.025 & 8.46 & 4.69 & 0.28 & -0.47 & 49 & 57 & 11 & 10 \\
570119 & 222.13518 & -0.57531 & 0.043 & 9.56 & 5.51 & 0.48 & 0.43 & 108 & 134 & 24 & 20 \\
570174 & 222.60503 & -0.46932 & 0.042 & 9.81 & 9.83 & 0.29 & -0.76 & 129 & 104 & 23 & 20 \\
573586 & 129.12557 & -0.08624 & 0.052 & 10.03 & 4.42 & 0.24 & 0.027 & 151 & 148 & 25 & 22 \\
574008 & 131.02735 & -0.10350 & 0.051 & 10.18 & 6.14 & 0.47 & 0.56 & 168 & 144 & 28 & 24 \\
574029 & 131.07729 & -0.04921 & 0.051 & 10.04 & 3.17 & 0.23 & 0.56 & 152 & - & 27 & 24 \\
574193 & 134.43437 & -0.04481 & 0.044 & 8.73 & 5.02 & 0.18 & -0.63 & 60 & 123 & 19 & 17 \\
574572 & 136.33633 & -0.03700 & 0.019 & 8.76 & 0.79 & 0.08 & -1.6 & 61 & - & 17 & 17 \\
574617 & 136.43827 & -0.19325 & 0.076 & 10.33 & 4.12 & 0.15 & 0.46 & 187 & - & 24 & 22 \\
574692 & 136.73747 & -0.12355 & 0.019 & 9.31 & 1.82 & 0.37 & -0.41 & 90 & 36 & 23 & 20 \\
583443 & 174.88168 & -0.15990 & 0.028 & 8.93 & 1.99 & 0.15 & -0.92 & 69 & 97 & 22 & 20 \\
583637 & 175.82494 & -0.18161 & 0.056 & 10.01 & 4.10 & 0.05 & 0.11 & 149 & - & 23 & 22 \\
584013 & 177.87898 & -0.07776 & 0.048 & 10.46 & 3.79 & 0.29 & 0.75 & 205 & 85 & 34 & 30 \\
585121 & 181.19288 & -0.01538 & 0.040 & 9.55 & 2.68 & 0.27 & -0.42 & 107 & 139 & 31 & 28 \\
585231 & 181.78147 & -0.02019 & 0.021 & 8.76 & 2.15 & 0.41 & -1.6 & 61 & 72 & 17 & 15 \\
592863 & 214.33856 & -0.16910 & 0.044 & 9.46 & 6.05 & 0.14 & -0.37 & 101 & 98 & 20 & 19 \\
592999 & 215.06156 & -0.07938 & 0.053 & 10.26 & 4.92 & 0.47 & 0.3 & 178 & 197 & 27 & 23 \\
593526 & 216.81878 & -0.09008 & 0.031 & 9.32 & 1.64 & 0.49 & -0.55 & 91 & - & 24 & 20 \\
594059 & 218.90933 & -0.09702 & 0.029 & 9.48 & 4.95 & 0.31 & -0.87 & 102 & 96 & 21 & 19 \\
594906 & 222.36208 & -0.16420 & 0.041 & 9.77 & 1.66 & 0.31 & 0.22 & 126 & - & 26 & 23 \\
594990 & 222.80149 & -0.06085 & 0.044 & 10.34 & 3.72 & 0.26 & -2.1 & 189 & 191 & 30 & 27 \\
598911 & 129.30130 & 0.38743 & 0.042 & 9.39 & 6.87 & 0.21 & -0.27 & 96 & 81 & 19 & 17 \\
598968 & 129.56040 & 0.35208 & 0.042 & 10.06 & 6.74 & 0.20 & 0.12 & 154 & 121 & 19 & 18 \\
599095 & 130.13599 & 0.26201 & 0.035 & 9.44 & 2.72 & 0.34 & -1.0 & 99 & 115 & 23 & 20 \\
599134 & 130.26050 & 0.39590 & 0.037 & 9.09 & 2.21 & 0.38 & -0.63 & 77 & - & 25 & 22 \\
599329 & 131.10371 & 0.34289 & 0.015 & 8.40 & 0.71 & 0.33 & -1.1 & 47 & - & 29 & 25 \\
599862 & 132.74012 & 0.23892 & 0.041 & 9.04 & 4.62 & 0.14 & -0.83 & 75 & 123 & 12 & 11 \\
600026 & 133.48520 & 0.21557 & 0.051 & 10.28 & 5.00 & 0.28 & 0.46 & 181 & 203 & 30 & 26 \\
600312 & 134.81541 & 0.39164 & 0.011 & 8.86 & 0.76 & 0.25 & -0.97 & 66 & 55 & 27 & 24 \\
601323 & 139.34146 & 0.32191 & 0.054 & 10.73 & 8.32 & 0.31 & 0.32 & 249 & 130 & 30 & 27 \\
601395 & 139.56851 & 0.38503 & 0.017 & 8.91 & 8.81 & 0.26 & -1.4 & 68 & 2 & 29 & 26 \\
610474 & 180.39356 & 0.34748 & 0.039 & 10.01 & 2.40 & 0.24 & 0.45 & 149 & - & 30 & 27 \\
610997 & 182.86904 & 0.37865 & 0.020 & 9.32 & 2.55 & 0.22 & -0.83 & 91 & 118 & 25 & 23 \\
611629 & 185.50338 & 0.31504 & 0.034 & 9.46 & 1.38 & 0.33 & -0.5 & 101 & - & 26 & 23 \\
617655 & 212.63506 & 0.22418 & 0.029 & 9.07 & 3.24 & 0.14 & -2.4 & 76 & 104 & 17 & 16 \\
617945 & 213.72345 & 0.40730 & 0.028 & 8.47 & 0.91 & 0.22 & -1.0 & 50 & - & 21 & 19 \\
618071 & 214.01854 & 0.21626 & 0.026 & 8.94 & 5.14 & 0.37 & -0.8 & 69 & 62 & 21 & 18 \\
618116 & 214.40555 & 0.32910 & 0.051 & 10.25 & 6.47 & 0.27 & 0.37 & 177 & 179 & 27 & 24 \\
618152 & 214.52287 & 0.22739 & 0.053 & 10.01 & 4.18 & 0.29 & -0.15 & 149 & 32 & 23 & 20 \\
619095 & 218.03502 & 0.41114 & 0.053 & 10.47 & 3.69 & 0.32 & 0.6 & 207 & 211 & 31 & 27 \\
622333 & 132.56179 & 0.75988 & 0.043 & 9.03 & 3.61 & 0.24 & - & 74 & 42 & 18 & 17 \\
622394 & 133.06978 & 0.68110 & 0.041 & 9.22 & 3.17 & 0.28 & -0.57 & 85 & 173 & 24 & 22 \\
622744 & 134.82995 & 0.79776 & 0.013 & 9.16 & 1.58 & 0.46 & -0.43 & 81 & 65 & 28 & 24 \\
622770 & 134.98662 & 0.78816 & 0.052 & 10.01 & 2.36 & 0.48 & -0.4 & 149 & - & 34 & 29 \\
623366 & 138.54711 & 0.81821 & 0.055 & 10.42 & 6.71 & 0.06 & 0.26 & 200 & - & 20 & 19 \\
623712 & 140.13867 & 0.72106 & 0.017 & 9.16 & 2.83 & 0.08 & -1.4 & 81 & - & 19 & 18 \\
\end{longtable}
\twocolumn
\clearpage



\bsp	
\label{lastpage}
\end{document}